\documentclass[]{spie}  %>>> use for US letter paper
\newcommand{\minus}{\scalebox{0.75}[1.0]{$-$}}

\usepackage{amsmath,amsfonts,amssymb}
\usepackage{graphicx}
\usepackage[colorlinks=true, allcolors=blue]{hyperref}
\usepackage{comment}

\usepackage{comment}
\usepackage{schemabloc}
\usepackage{mathtools}

\usepackage{subcaption}

\title{Development of the ComPair gamma-ray telescope prototype}

\author[a]{Daniel Shy}
\author[b]{Carolyn Kierans}
\author[b,c,d]{Nicholas Cannady}
\author[b]{Regina Caputo}
\author[e]{Sean Griffin}
\author[f]{J. Eric Grove}
\author[b]{Elizabeth Hays}
\author[g]{Emily Kong}
\author[h]{Nicholas Kirschner}
\author[b,i]{Iker Liceaga-Indart}
\author[b]{Julie McEnery}
\author[b]{John Mitchell}
\author[b,j,d]{A. A. Moiseev}
\author[k]{Lucas Parker}
\author[b]{Jeremy S. Perkins}
\author[f]{Bernard Phlips}
\author[b,j,d]{Makoto Sasaki}
\author[b,m]{Adam J. Schoenwald}
\author[f]{Clio Sleator}
\author[l]{Jacob Smith}
\author[j]{Lucas D. Smith}
\author[i,b,d]{Sambid Wasti}
\author[f]{Richard Woolf}
\author[f]{Eric Wulf}
\author[b,c,d]{Anna Zajczyk}

\affil[a]{National Research Council Research Associate resident at the U.S. Naval Research Laboratory, 4555 Overlook Ave., SW, Washington, DC, 20375, USA}
\affil[b]{NASA Goddard Space Flight Center, Greenbelt, MD, USA}
\affil[c]{Center for Space Sciences and Technology, University of Maryland, Baltimore County, 1000 Hilltop Circle, Baltimore, MD 21250, USA}
\affil[d]{Center for Research and Exploration in Space Science and Technology, NASA/GSFC, Green-
belt, MD 20771, USA}
\affil[e]{Wisconsin IceCube Particle Astrophysics Center, University of Wisconsin-Madison, 222 W Washington Ave Unit 500, Madison, WI 53703, USA}
\affil[f]{Space Science Division, Naval Research Laboratory, 4555 Overlook Ave SW, Washington, DC
20375}
\affil[g]{Technology Service Corporation, Arlington, VA, 22202, USA}
\affil[h]{The Department of Physics, The George Washington University, 725 21st NW, Washington, DC 20052, USA}
\affil[i]{Catholic University of America, 620 Michigan Ave NE, Washington, DC 20064, USA}
\affil[j]{University of Maryland at College Park, College Park, MD 20742, USA}
\affil[k]{Los Alamos National Laboratory, Los Alamos, NM 87544, USA}
\affil[l]{George Mason University, resident at the Naval Research Laboratory, 4555 Overlook Ave SW, Washington, DC 20375}
\affil[m]{University of Maryland, Baltimore County, 1000 Hilltop Circle, Baltimore, MD 21250, USA}

\authorinfo{Corresponding author: D. Shy, e-mail: daniel.shy.ctr@nrl.navy.mil \\ DISTRIBUTION A: Approved for public release, distribution is unlimited.}

\begin{document} 
\maketitle

\begin{abstract}
There is a growing interest in the science uniquely enabled by observations in the MeV range, particularly in light of multi-messenger astrophysics. The Compton Pair (ComPair) telescope, a prototype of the AMEGO Probe-class concept, consists of four subsystems that together detect and characterize gamma rays in the MeV regime. A double-sided strip silicon Tracker gives a precise measure of the first Compton scatter interaction and tracks pair-conversion products. A novel cadmium zinc telluride (CZT) detector with excellent position and energy resolution beneath the Tracker detects the Compton-scattered photons. A thick cesium iodide (CsI) calorimeter contains the high-energy Compton and pair events. The instrument is surrounded by a plastic anti-coincidence (ACD) detector to veto the cosmic-ray background. In this work, we will give an overview of the science motivation and a description of the prototype development and performance.
\end{abstract}


\section{INTRODUCTION}
\label{sec:intro}

Gamma rays have an integral part in multi-messenger discoveries such as the combined detection with neutrinos from supernova
SN1987A~\cite{SupernovaGammaRays, supernovaGamma} and gravitational waves (GWs) from GW 170817~\cite{GW170817}. However, the MeV domain remains largely under explored; COMPTEL on the Compton Gamma Ray Observatory was the last major imaging instrument sensitive to that energy range and was de-orbited in 2000~\cite{ASTROMEV}.

The All-sky Medium Energy Gamma-ray Observatory (AMEGO) is a probe-class mission concept designed to address the MeV gap~\cite{amegoJulie}. AMEGO is sensitive to photons from $0.2 \ \mathrm{MeV}$ to $10 \ \mathrm{GeV}$, which covers both the Compton and pair-conversion regime. To accomplish the detection and imaging of both these processes, we employ a unique combination of a `Tracker' with both a low- and high-energy calorimeter that are all surrounded by an anti-coincidence detector. These four major subsystems are briefly summarized here:

\begin{itemize}
    \item \textbf{Double Sided Silicon Strip Detector (DSSD) Tracker} acts as the Compton-scattering element for low-energy gamma rays and the pair-conversion material for high-energy gamma rays by measuring the position and energy of the particles moving through the Tracker layers.
    \item \textbf{Cadmium Zinc Telluride (CZT) calorimeter}, or low-energy calorimeter, is optimized for detecting the scattered photon and enhances the line sensitivity, as well as providing improved imaging and polarization capabilities.
    \item \textbf{Cesium Iodide (CsI) calorimeter}, or high-energy calorimeter, is designed to extend the energy range by detecting the ensuing electromagnetic showers. 
    \item \textbf{Anti-coincidence detector (ACD)} consisting of a plastic scintillator to reject charged particle background.
\end{itemize}

\noindent The simulated performance of the AMEGO concept is available in Ref.~\citenum{Kierans_Amego} which is optimized for continuum sensitivity across a broad energy range. There are other astrophysical instruments with a slightly different mission set and technological solution in various stages of conception/development. They include the Compton Spectrometer and Imager (COSI)~\cite{COSI_SMEX}, ASTROGAM~\cite{ASTROMEV}, Galactic Explorer with a Coded Aperture Mask Compton Telescope (GECCO), Advanced Particle-astrophysics Telescope (APT)~\cite{APT}, SMILE-2+~\cite{SMILE}, and AMEGO-X~\cite{Caputo2022AMEGO, amegox}.

To demonstrate the instrument's potential, the AMEGO team has developed a prototype, ComPair, to advance the hardware and software used in the probe concept. Like the AMEGO probe, it consists of all four subsystems but with an energy range of $0.5-100 \ \mathrm{MeV}$. After in-lab calibration and qualification, the prototype is slated to perform a balloon test flight.

This work presents an update on the development and initial performance of the ComPair instrument. Sec.~\ref{sec:instrument} presents an overview of the ComPair prototype while Sec.~\ref{sec:performance} presents the preliminary performance of each subsystem as well as the integrated instrument. The manuscript also describes the planned future work.

\begin{figure}[h]
    \centering
    \subfloat[Front CAD view of ComPair]{%
    \includegraphics[trim={0 0 0cm 0cm},clip,height=0.31\textheight]{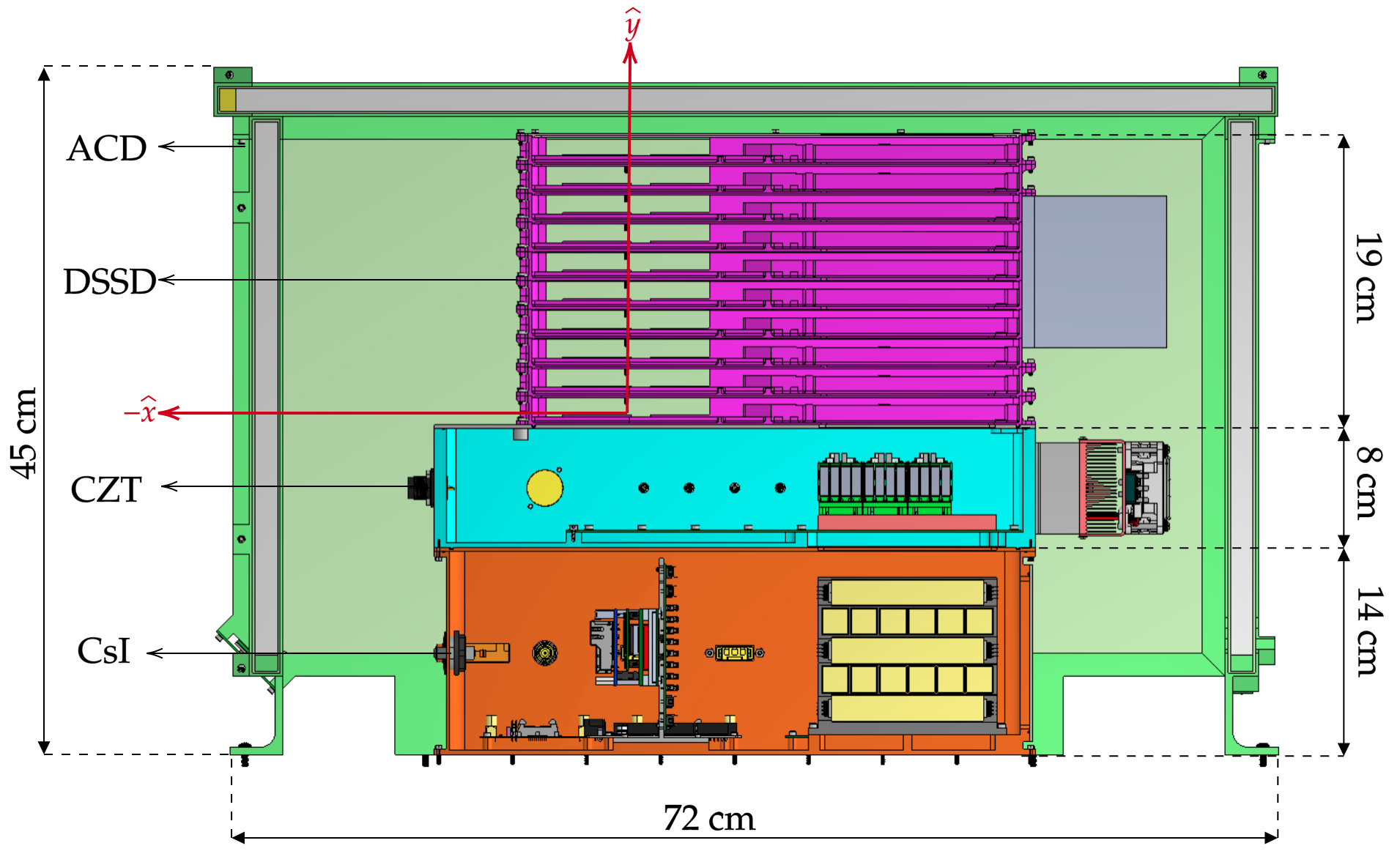}}
    \hspace*{0.2cm}
    \subfloat[Setup of the ComPair prototype without the ACD]{%
    \includegraphics[trim={0 0 20cm 0cm},clip, angle=-90,origin=c, height=0.27\textheight]{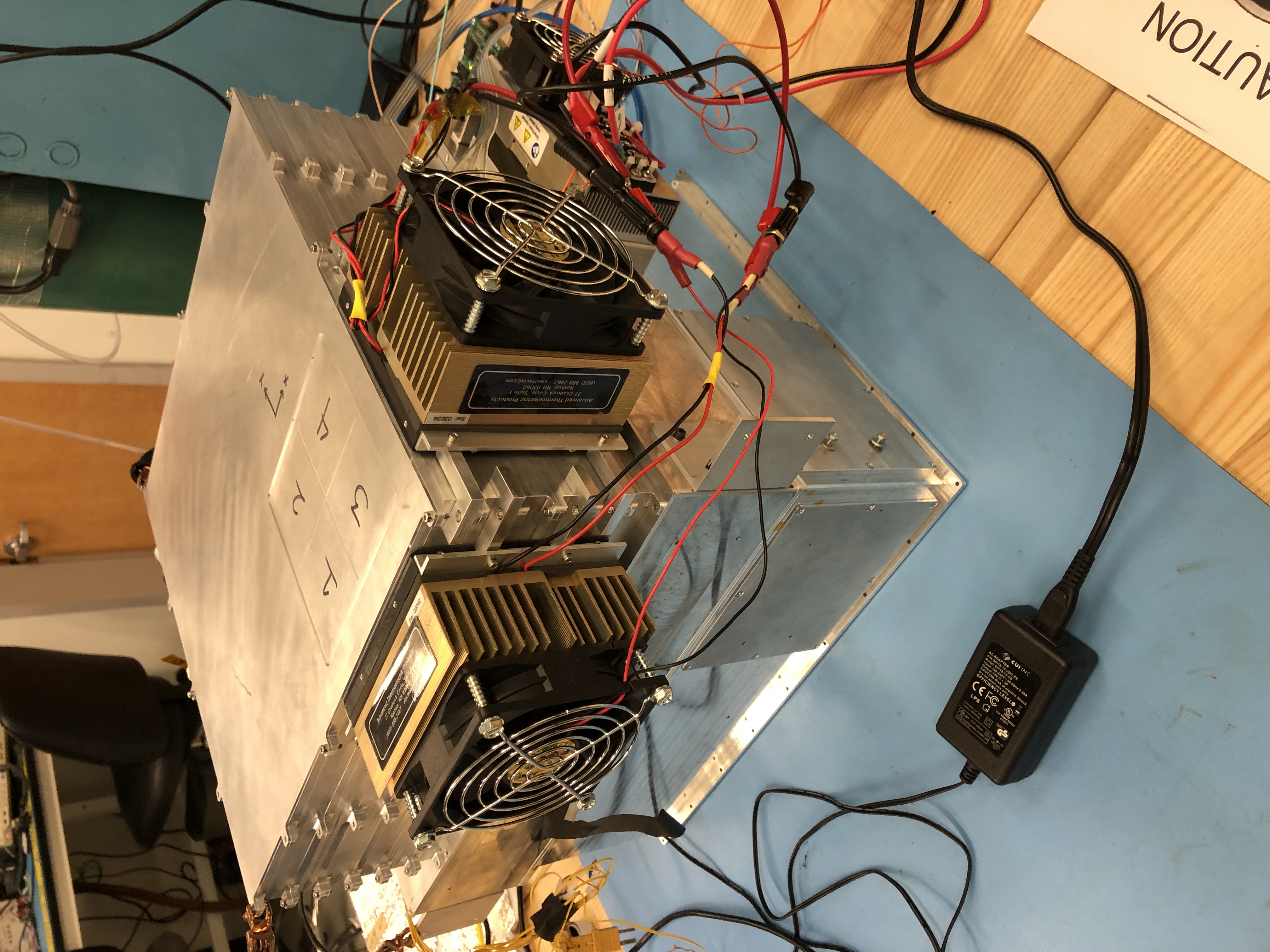}}
\caption{(a) Cutaway view of the CAD showing ComPair's subsystems and their dimensions. (b) Image of the ComPair prototype, without the ACD, integrated with 6 Tracker layers, CZT, and the CsI calorimeters. Two peltier coolers are attached to the Tracker layers for heat removal. }
\label{fig:CAD}
\end{figure}

%\begin{comment}

\section{ComPair: An AMEGO Prototype}
\label{sec:instrument}

Fig.~\ref{fig:CAD}a presents a computer-aided design (CAD) model of the ComPair balloon prototype, while Fig.~\ref{fig:CAD}b presents the actual experimental setup without the ACD. The three core subsystems (Tracker, CZT and CsI) all fit within an envelope of roughly $43 \times 30 \times 41 \ \mathrm{cm}^3$. The DSSD Tracker layers are at the top of the instrument. Below the Tracker are arrays of the CZT detectors, and the CsI calorimeter is at the bottom of the instrument. 

For the actual balloon flight, the CZT housing will be re-designed to be hermetically sealed. The entire balloon instrument with the associated electronics weighs less than $100 \ \mathrm{kg}$. Additional information on the balloon instrument is available in Sec.~\ref{sec:futureWork}. The rest of this section overviews the different subsystems, trigger module, and flight computer.

\subsection{Tracker: Double Sided Silicon Strip Detector}

The ComPair Tracker has been designed, built, and tested by NASA's Goddard Space Flight Center (GSFC), with firmware support from Los Alamos National Laboratory (LANL). Its goal is to achieve position sensitivity within each Tracker layer to measure the Compton-scattered electron or the $e^- \minus e^+$ tracks from the pair production. During a Compton scatter, should the recoiled electron proliferate into other layers, the Tracker also has the potential to track the electron and produce Compton arcs rather than rings~\cite{mega}.

\begin{figure}[h]
    \centering
    \subfloat[A single layer of the Tracker. The DSSD is in the custom carrier on the lower right with 6 ASICs placed right above it while the remaining 6 are placed on the posterior side of the board. The digital FPGA board is on the top left.]{%
    \includegraphics[height=0.25\textheight]{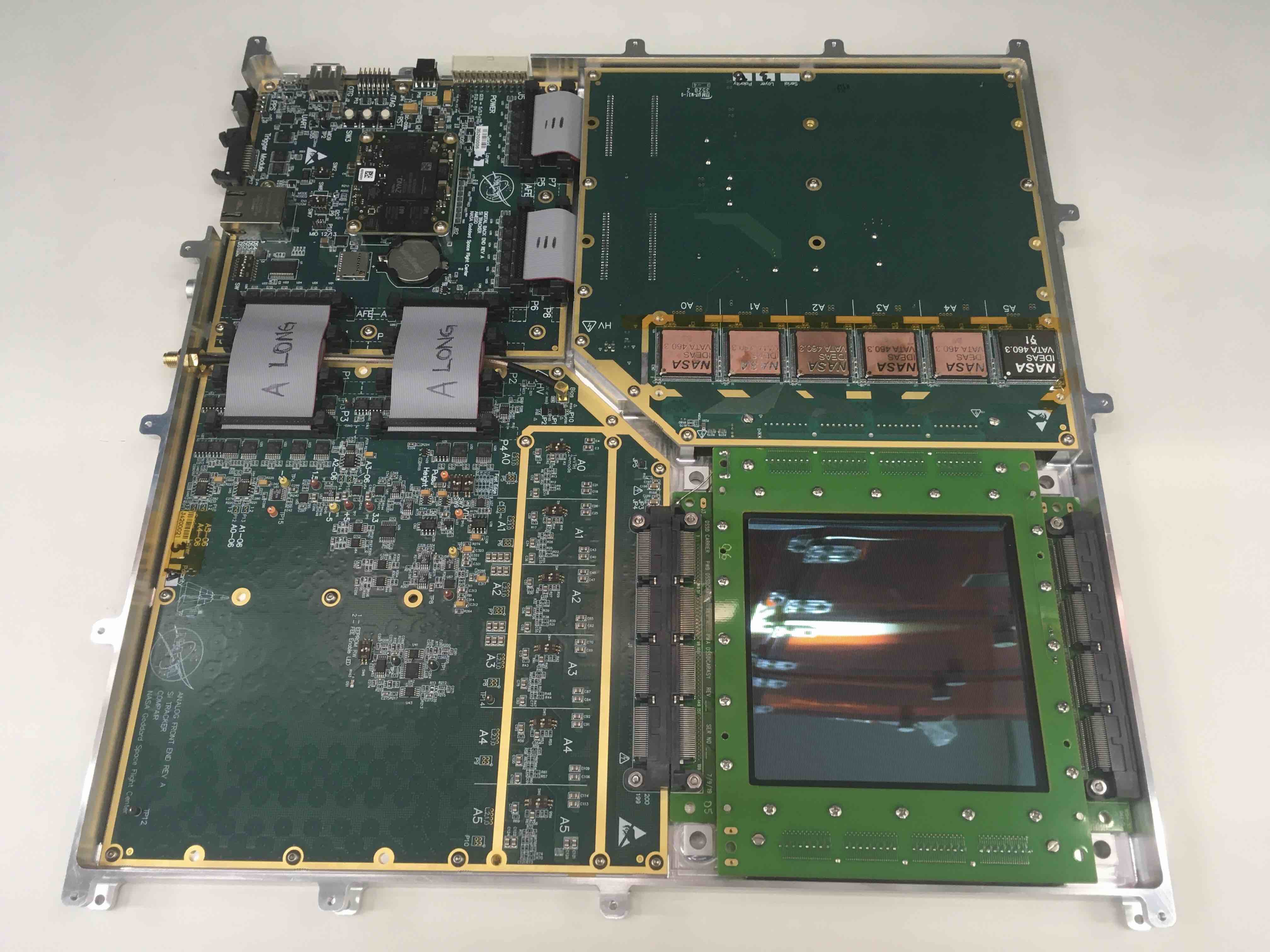}}
    \hspace*{1 cm}
    \subfloat[The Tracker integrated with 6 of its total of 10 layers for testing.]{%
    \includegraphics[height=0.25\textheight]{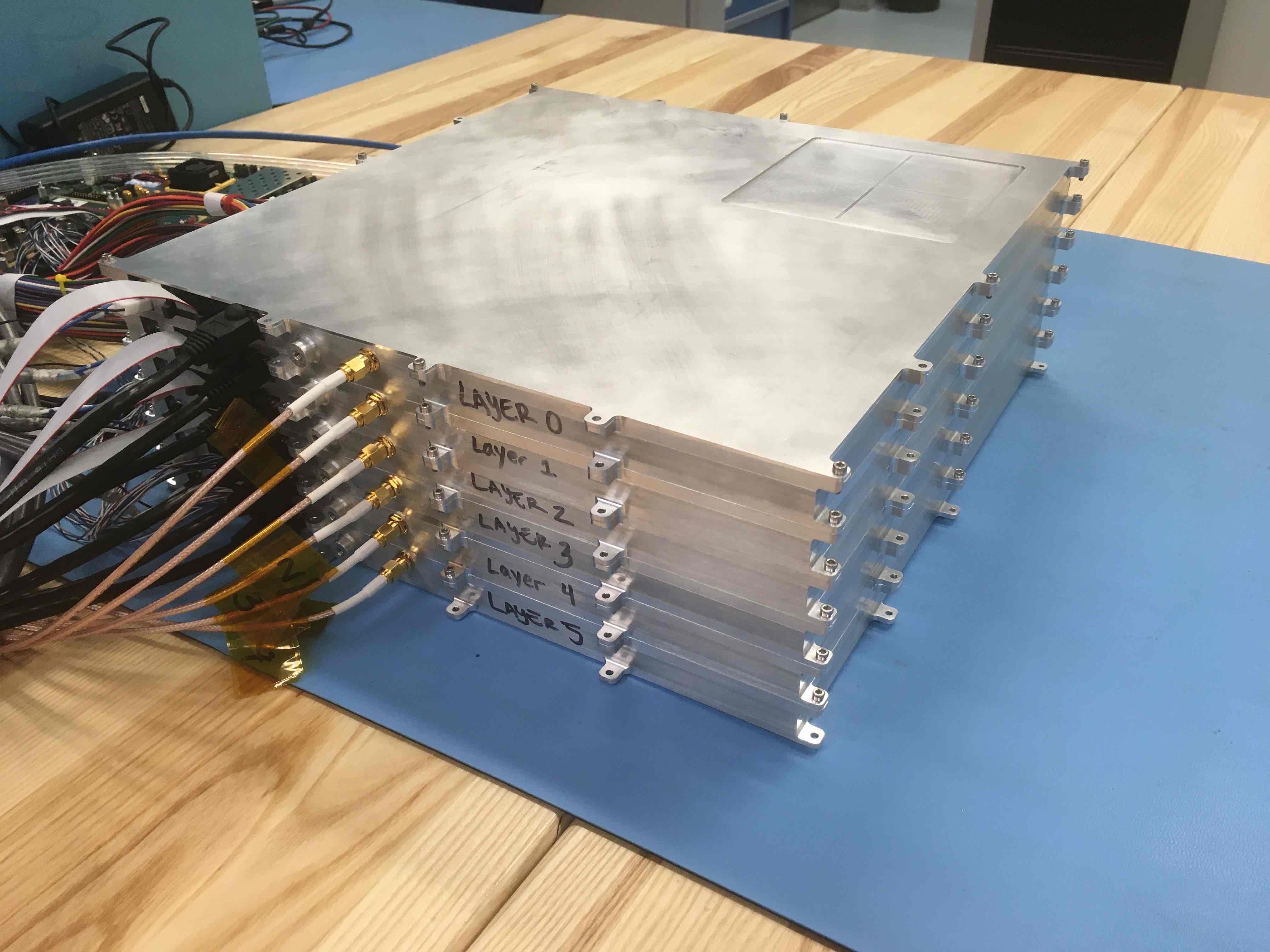}}
\caption{The ComPair DSSD Tracker subsystem with the custom analog and digital readouts.}
\label{fig:dssd}
\end{figure}

Fig.~\ref{fig:dssd}a presents a single layer of the Tracker. The DSSD is positioned on the lower-right side of the image and has an active area of $10 \times 10 \ \mathrm{cm}^2$. Fabricated by Micron Semiconductor Ltd, the silicon wafers are $500 \ \mu\mathrm{m}$ thick with strips on each side that are orthogonal to each other to reconstruct the $X$ and $Y$ positions of interaction~\cite{trackDevIEEE}. The strips are $60 \ \mu\mathrm{m}$ wide with a pitch of $\sim 510 \ \mu\mathrm{m}$. This results in 192 AC-coupled strips per side.

The wafers are mounted in a custom carrier that routes the signals to front-end electronics that use the IDEAS VATA460.3 application-specific integrated circuits (ASIC)~\cite{ideasVata460}. Each side of the DSSD requires 6 ASICs with 32 channels on each ASIC. A complete description of the readout is available in Ref.~\citenum{trackerDevCompair}. The ComPair prototype aims to use ten layers. Fig.~\ref{fig:dssd}b presents 6 fully integrated layers.

\subsection{Low-Energy Calorimeter: CZT}
\begin{figure}[h]
\centering
    \subfloat[Single CZT bar]{%
    \includegraphics[trim={1cm 15cm 1cm 20cm},clip,height=0.21\textheight]{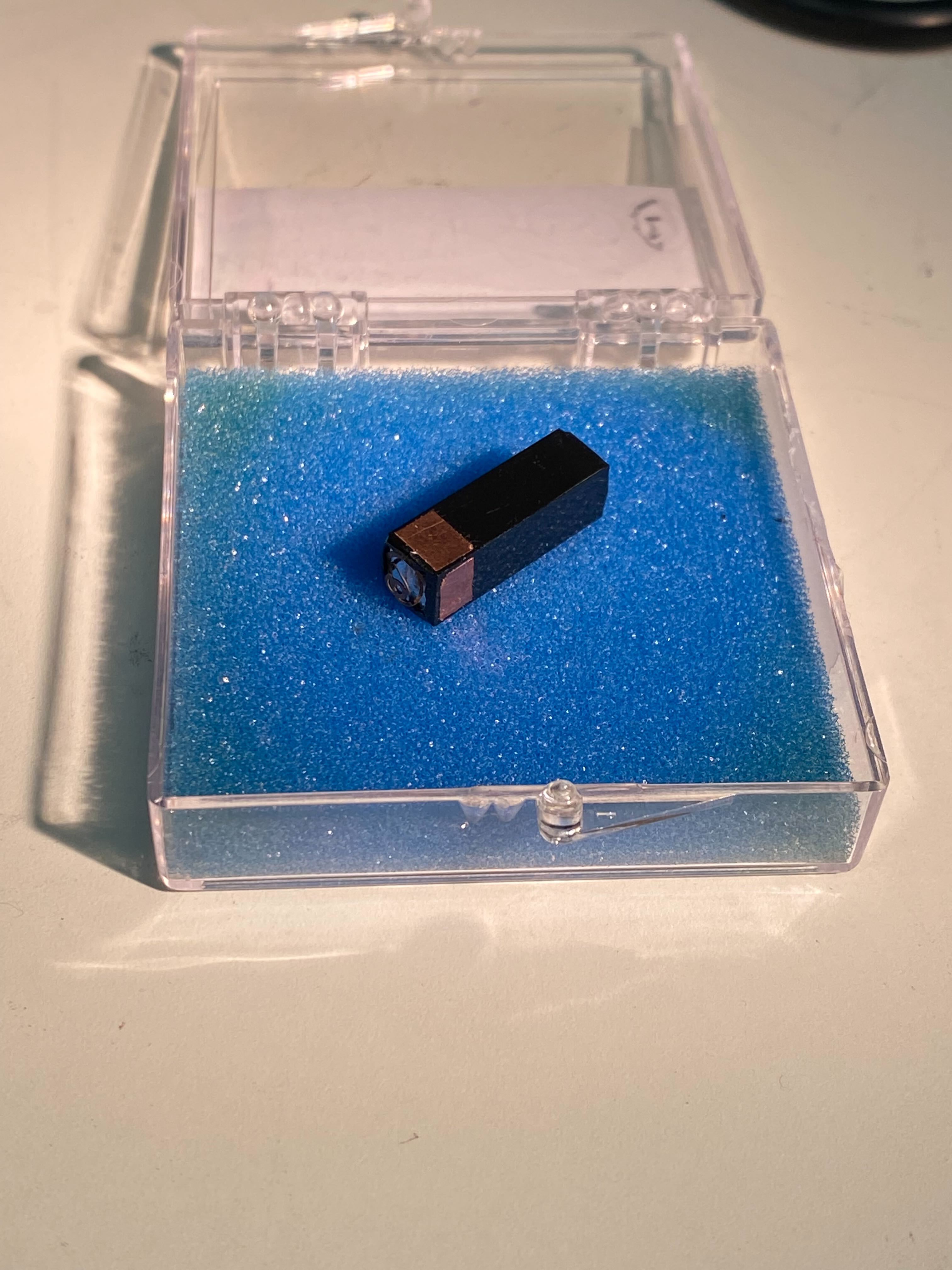}}
    \hspace*{0.5 cm}
    \subfloat[Single CZT crate]{%
    \includegraphics[height=0.21\textheight]{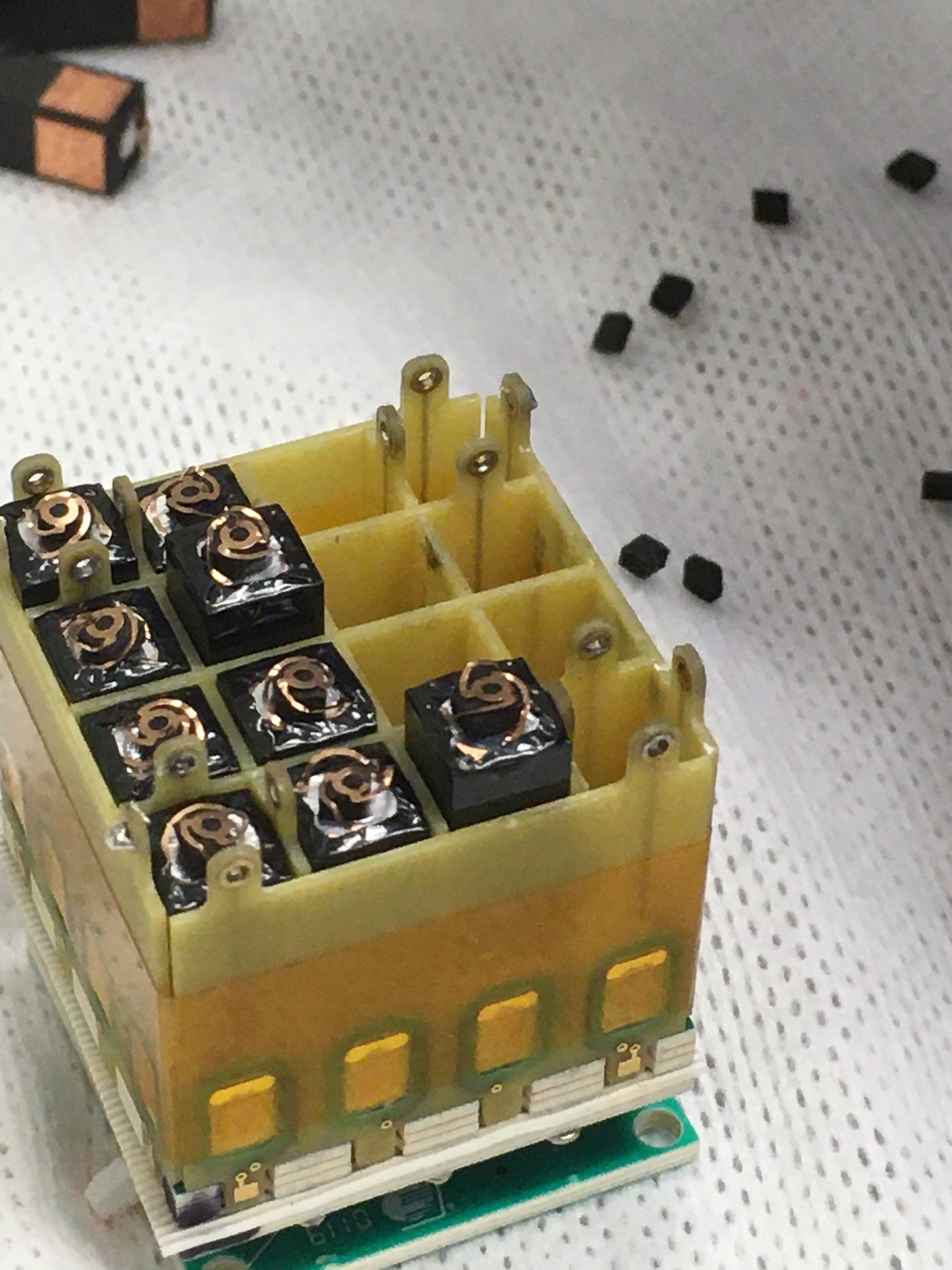}}
    \hspace*{0.5 cm}
    \subfloat[9 crates as implement in the ComPair system]{%
    \includegraphics[height=0.21\textheight]{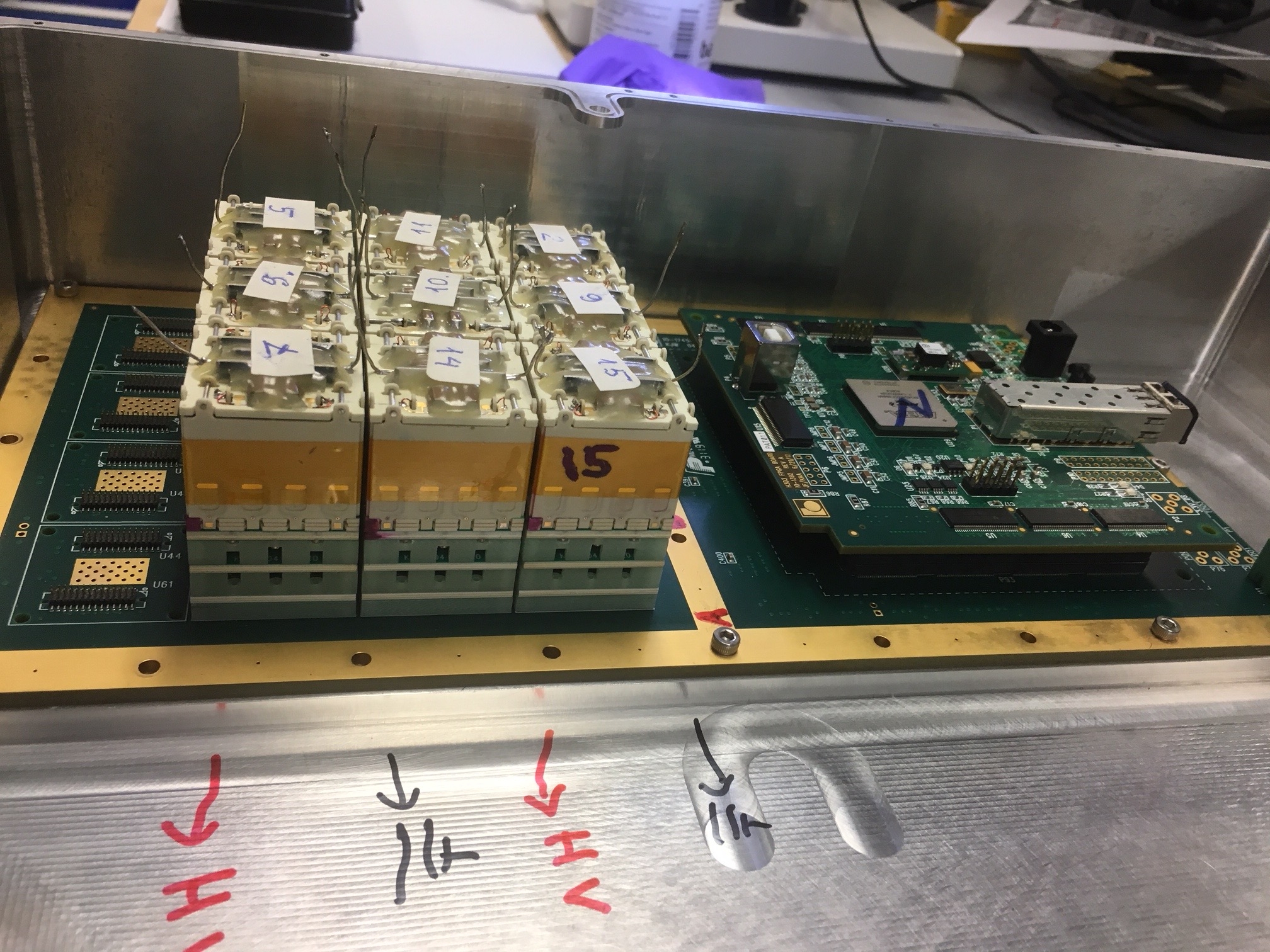}}
\caption{Images of the CZT Subsystem. (a) Shows a single CZT bar with the pads that create the virtual Frisch grid clear visible at the edge of the crystal. (b) A semi-populated CZT crate. (c) The assembled CZT subsystem with the FPGA board is visible on the right side of the image.}
\label{fig:cztSubsystem}
\end{figure}

The low-energy calorimeter uses 3D position-sensitive virtual Frisch-grid Cadmium Zinc Telluride (CZT) developed at Brookhaven National Laboratory~\cite{bnlCZT, FrischGridCZT}. Its performance is optimized to act as the absorber plane for gamma rays scattered in the Tracker. Its major advantage over other semiconductors is that it can operate at room temperature due to its wide bandgap. Each CZT crystal, manufactured by Redlen Technologies, is $6 \times 6 \times 20 \ \mathrm{mm}^3$ in size. Fig.~\ref{fig:cztSubsystem}a illustrates a single CZT crystal. Each end has a gold electrode representing the cathode and anode. Due to the severe trapping of the holes in CZT, single polarity charge sensing techniques need to be applied~\cite{ShockleyRamoHe}. In this work, we utilize a virtual Frisch-grid technique whereby the shielding electrode pads are placed on the sides of the crystal. This allows for the reconstruction of the 3D position of interaction.

First, the cathode-to-anode ratio (CAR) calculates the depth of interaction ($Z$ coordinate). Due to the single-polar nature of CZT, the signal on the cathode is depth-dependent, while the virtual Frisch grid allows for an induced charge on the anode that is depth-independent. Therefore, the ratio results in a relative depth of interaction. Using the technique, the depth resolution is estimated to be less than 0.5 mm. Next, the side pads and the relative amplitudes induced on each pad could be used to calculate the $X\minus Y$ coordinates of interaction~\cite{4x4CZT}. Therefore, assuming a linear approximation for the response function, the $X\minus Y$ coordinate could be calculated as

\begin{equation}\label{eq:cztSubBarCoordinate}
\begin{split}
X = \frac{A^x_1}{A^x_1 + A^x_2}, \\
Y = \frac{A^Y_1}{A^Y_1 + A^Y_2},
\end{split}
\end{equation}

\noindent where $A^x_1$ and $A^x_2$ represent the recorded amplitude of pad 1 that is opposite of pad 2 in the $X$ direction. $A^y_1$ and $A^y_2$ similarly represent the pads in the $Y$ direction. This method can produce a position resolution of better than $1 \ \mathrm{mm}$ in $X-Y$. Since the 3D position of each interaction can be reconstruction, a 3D voxel correction can be applied. This technique can compensate for non-uniformity in the detector material and results in a resolution of around 1\% FWHM at 662 keV.

As a subsystem for ComPair, each crystal is packaged in a crate that encloses a $4\times4$ array of crystals, shown in Fig.~\ref{fig:cztSubsystem}b. The crate is read out by a custom low-noise ASIC called the advanced virtual Frisch-Grid 2 (AVG2)~\cite{bnlAVG}. An Altera Arria GX FPGA then handles the data recording and ASIC controls. In ComPair, there are a total of 9 crates arranged in a $3\times 3$ array, shown in Fig.~\ref{fig:cztSubsystem}c. The bars are biased to 2500 volts which requires the need for it to be placed in a pressure vessel for the balloon flight. 

\subsection{High-Energy Calorimeter: CsI}
\begin{figure}[h]
\centering
    \subfloat[The bare SiPMs on left and SiPM mounted bars on right.]{%
    \includegraphics[angle=-90,origin=c, height=0.252\textheight]{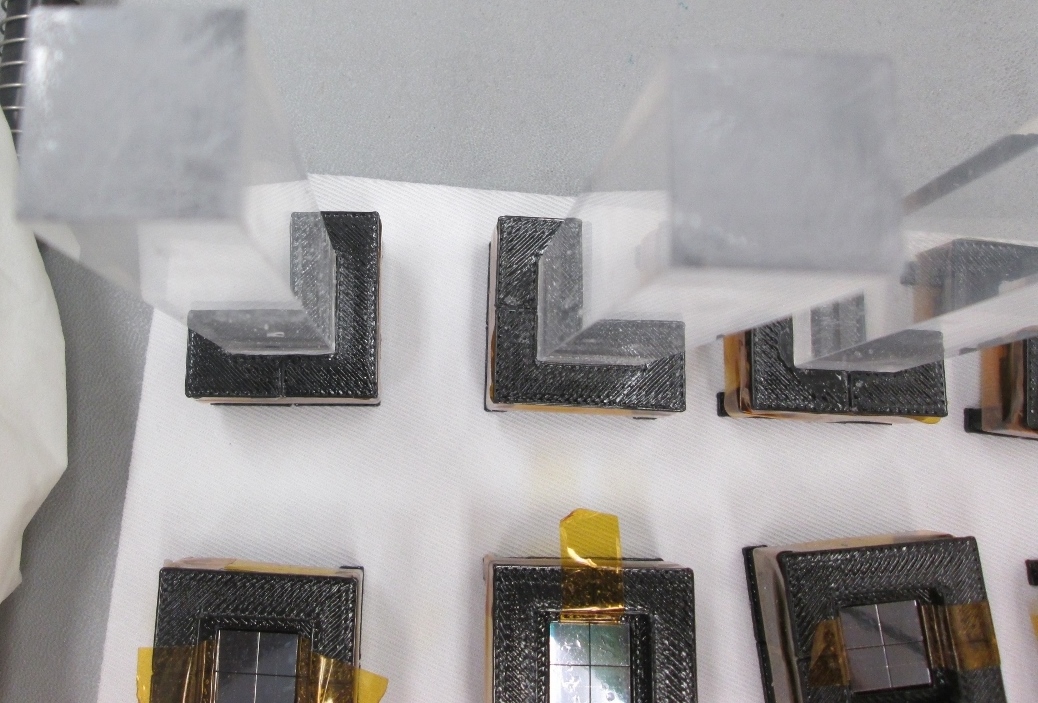}}
    \hspace*{1 cm}
    \subfloat[The CsI Calorimeter subsystem.]{%
    \includegraphics[trim={1.7cm 1.75cm 1.5cm 0.5cm},clip,height=0.3\textheight]{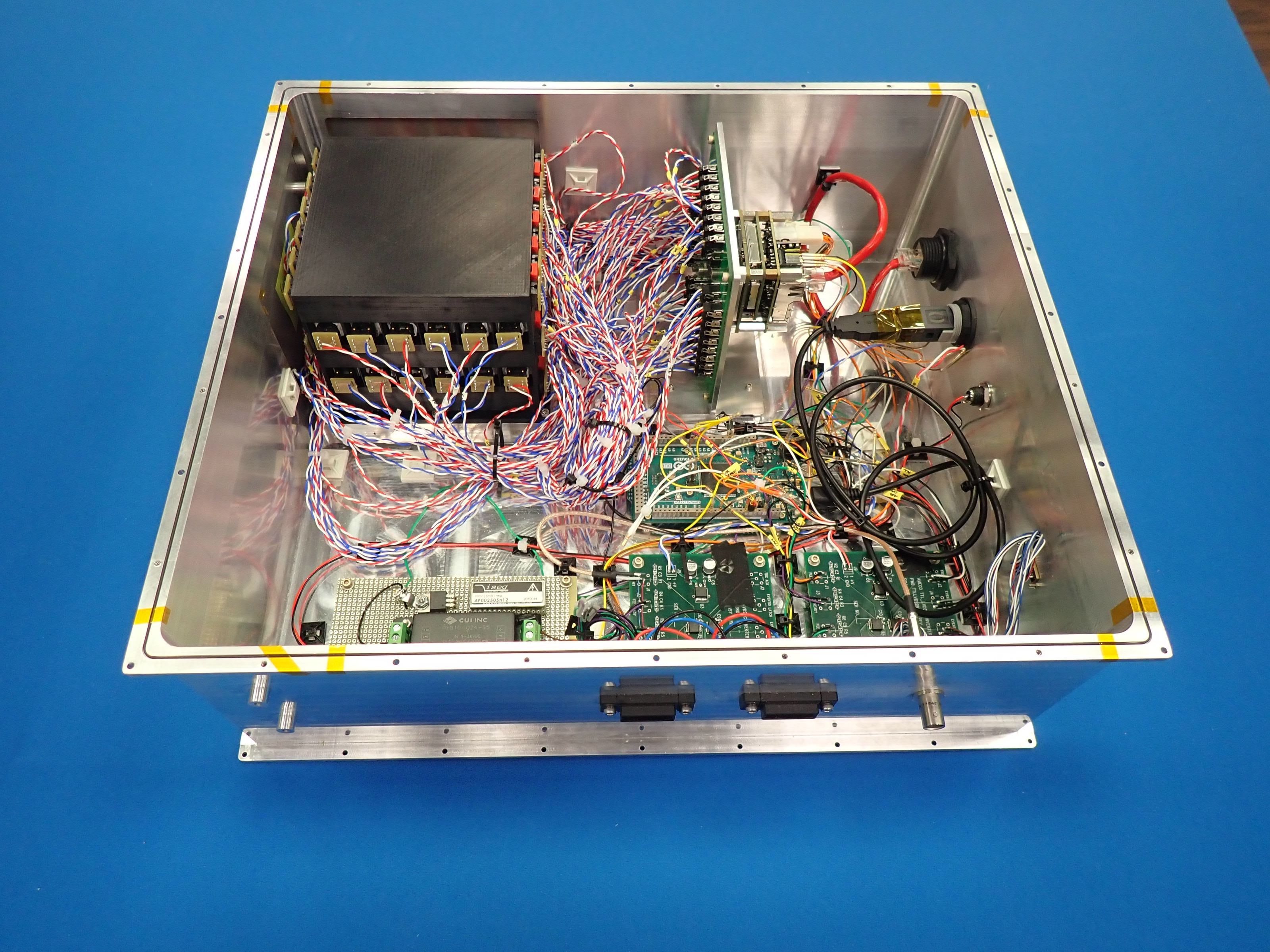}}
\caption{(a) The CsI bars at various stages of SiPM bonding. (b) The CsI calorimeter subsystem within its housing. The black assembly in the upper-left corner of the picture contains the arrays of CsI. Immediately to the right is the ROSSPAD front-end readout. Below are electronics associated with the readout and synchronization with the trigger module.}
\label{fig:subsystem_CSI}
\end{figure}

Developed by the U.S. Naval Research Laboratory, the high-energy calorimeter is composed of CsI:Tl bars and takes heritage from Fermi's Large Area Telescope (LAT) CsI:Tl Calorimeter~\cite{FermiCalorimeter}. Each bar, grown by either Amcrys in Kharkov, Ukraine or Saint-Gobain, has dimension $1.67 \times 1.67 \times 10.0 \ \mathrm{cm}^3$. Each face of the scintillator is roughened with 180-grit sandpaper to increase light asymmetry along the length of the bar. The crystals are then wrapped with Tetratex to increase light output~\cite{CsIDevComPair}. Each end of the bar is read out by a $2 \times 2$ array of $6 \times 6 \ \mathrm{mm}^2$ J-series SiPMs manufactured by ON Semiconductors. Fig.~\ref{fig:subsystem_CSI}a shows the SiPMs being bonded with epoxy to the crystal bars.

The CsI calorimeter makes use of 30 bars arranged in a hodoscopic fashion. In other words, the bars are arranged in rows of 6 that alternate direction between each layer. The bars are placed in a 3D printed plastic housing such that the center-to-center pitch between each bar is $1.9 \ \mathrm{cm}$. The 60 SiPMs are read out by an IDEAS ROSSPAD~\cite{rosspad}, which is a module with 64 channels that is distributed to four SiPHRAs ASICS~\cite{SIPHRA}. If $E_1$ and $E_2$ represent the energy recorded on each end of the bar, the depth of interaction in the bar is calculated as $\mathrm{DOI} = (E_1-E_2)/(E_1+E_2)$ and the energy deposited is calculated as $E_{\mathrm{dep}} = \sqrt{E_1\times E_2}$. The gain and SiPM voltage is tuned such that each bar has a dynamic energy range of roughly 250 keV to 30 MeV. Fig.~\ref{fig:subsystem_CSI}b shows the CsI calorimeter in its final configuration.

\subsection{ACD}
\begin{figure}[h]
\centering
    \subfloat[Exploded CAD view of a panel]{%
    \includegraphics[trim={0cm 0.1cm 3.3cm 0cm},clip, height=0.21\textheight]{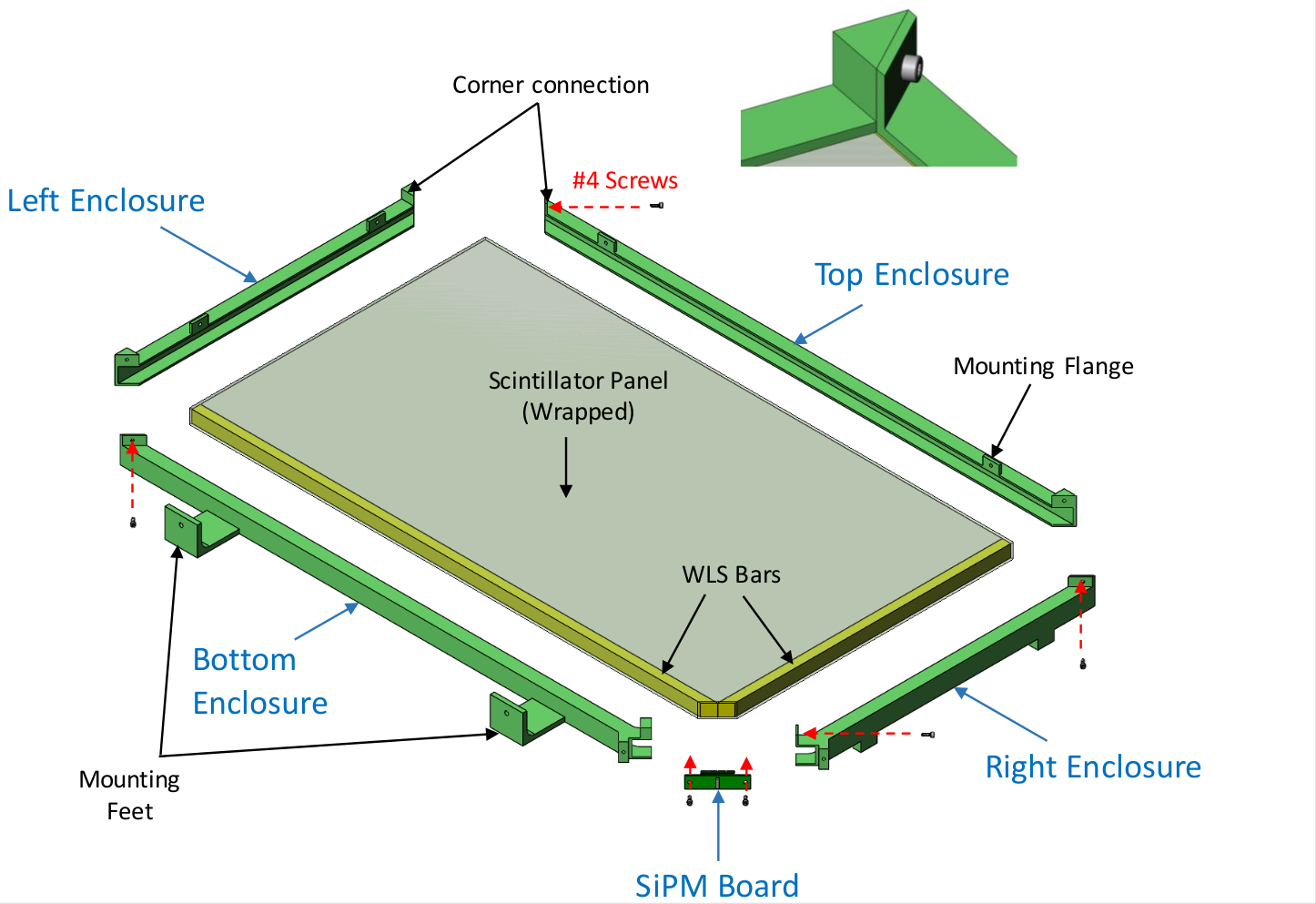}}
    \hspace*{0.15 cm}
    \subfloat[Construction of the ACD panel]{%
    \includegraphics[trim={0cm 0.0cm 0cm 0cm},clip, height=0.21\textheight]{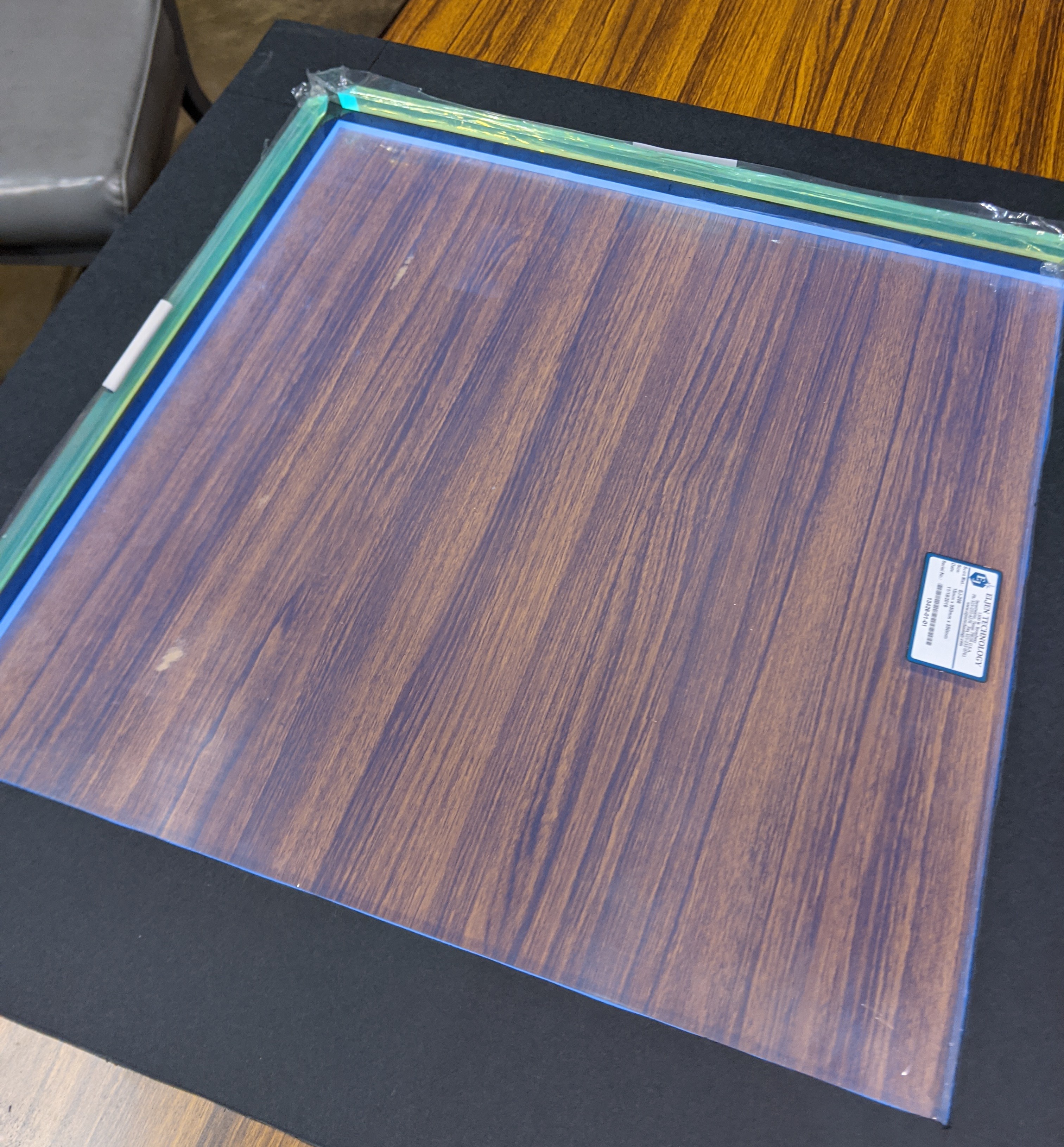}}
    \hspace*{0.15 cm}
    \subfloat[The ACD subsystem.]{%
    \includegraphics[trim={2cm 0.1cm 1.5cm 1.5cm},clip,height=0.21\textheight]{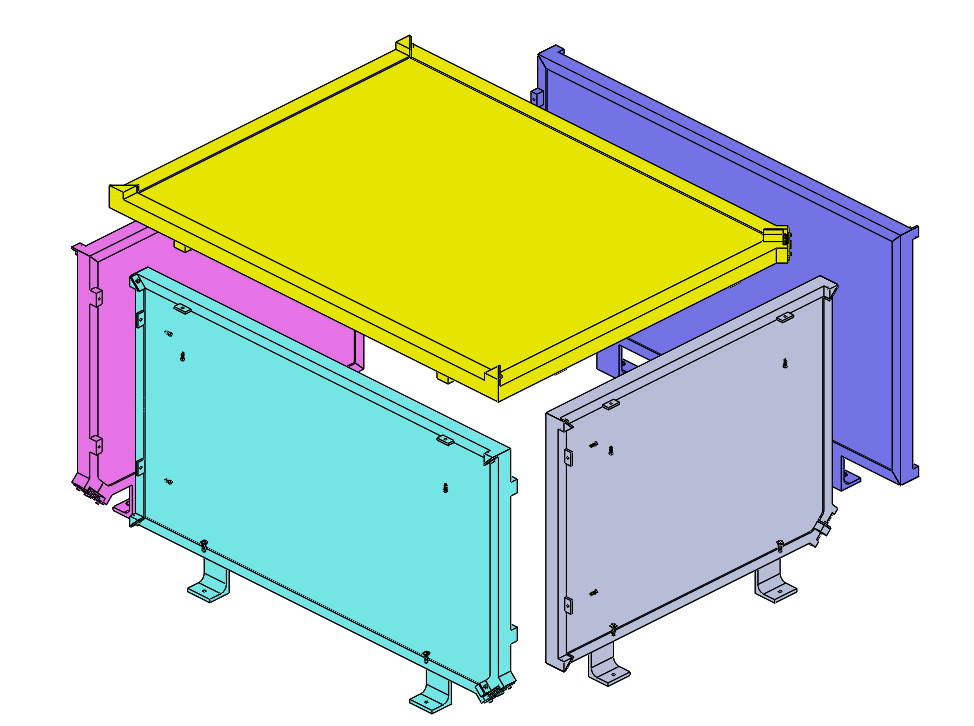}}
\caption{(a) The plastic scintillator panel is in the center with the two wavelength shifters shown in yellow. The readout edge and the SiPMs are shown at the bottom of the image. The panel is then enclosed by the aluminum frame (darker green). The inset image shows the joint between two frame enclosures. (b) Shows the assembly of the wavelength shifters onto the plastic panel. (c) Exploded CAD view of the 5 ACD panels as they would assemble in the final configuration.}
\label{fig:subsystem_ACD}
\end{figure}

Similarly to the Fermi-LAT, the ComPair ACD uses plastic scintillator panels, specifically Eljen EJ-208. Unlike the Fermi LAT, the panels are not tiled as there is less concern of self-vetoing at the lower AMEGO energies. Two wavelength shifters are appended to two adjacent edges, shown in Fig.~\ref{fig:subsystem_ACD}a. At the edge where the two shifters meet, the edges are chamfered out such that two $2\times 2$ arrays of $6\times6 \ \mathrm{mm^2}$ C-Series SiPMs by On Semiconductors can be placed and read out the scintillation signals. The panels are then placed in aluminum frames for structural support.

Five panels compose the entire subsystem and cover the instrument's four sides and its top. The ACD's dimensions span $72.1 \times 56.1 \times 45.2 \ \mathrm{cm^2}$, as seen in Fig.~\ref{fig:subsystem_ACD}b. The ACD is then mounted onto a base plate with a bracket that elevates the panel slightly such that harnessing can pass from the core instrument to outside the ACD to connect to their relevant electronics. The front-end readout is duplicated from the CsI calorimeter. 

\subsection{Trigger Module}

\begin{figure}[h!]
\includegraphics[trim={5cm 2cm 7cm 5cm},clip, angle=180,origin=c, width=0.6\linewidth]{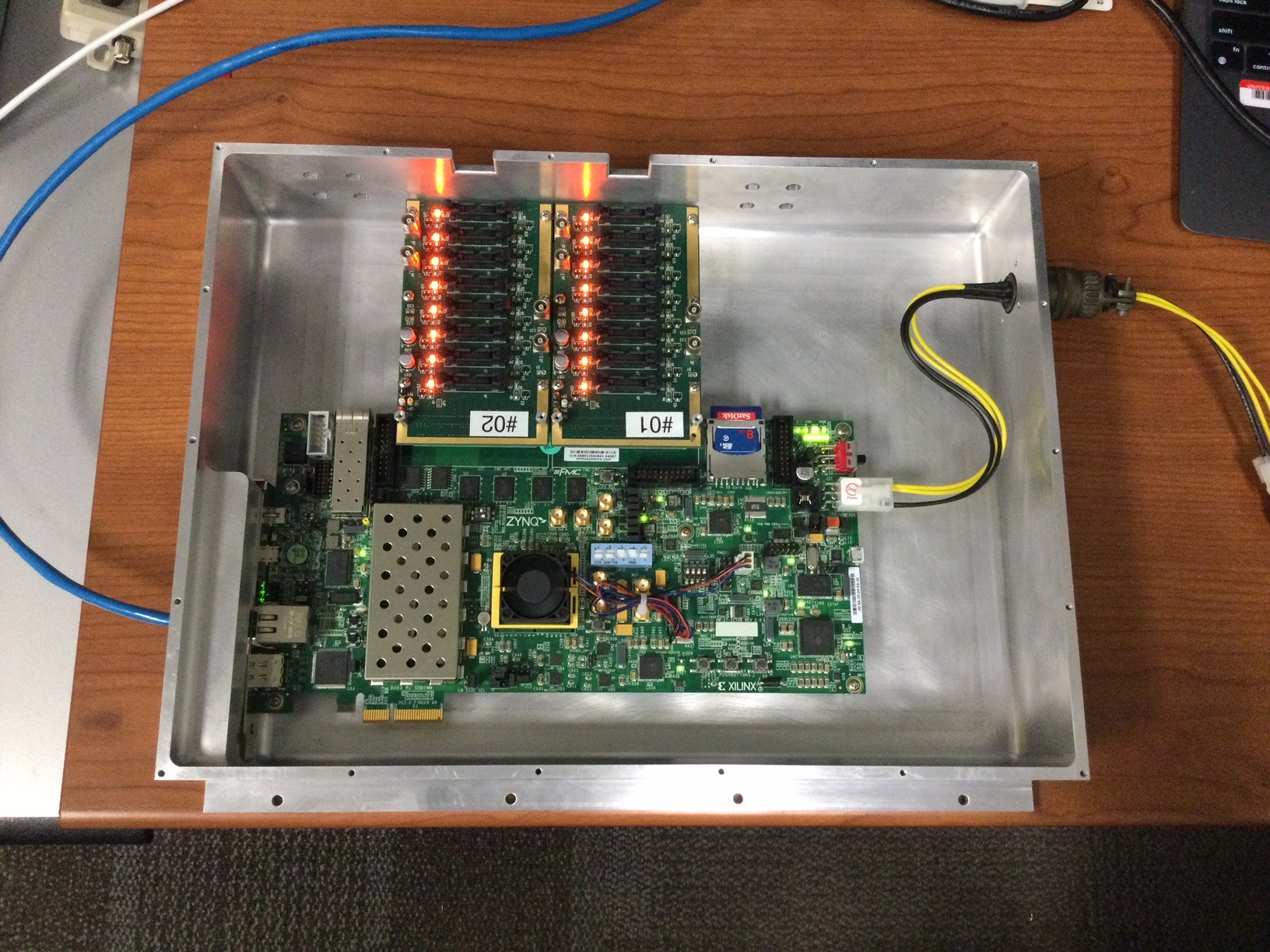}
  \centering
  \caption{Picture of the ComPair Trigger Module. This module utilizes the ZC706 Xilinx Zynq-7000 kit, which will be replaced with the ZC702 for the balloon flight.}
  \label{fig:triggerModule}
\end{figure}

The Trigger Module (TM) is designed and built at GSFC to check for coincidences between the subsystems within a programmable window~\cite{MakotoTM}. It uses a Xilinx Zynq-7000 FPGA evaluation kit with a custom breakout board for the 14 separate front-end boards: 10 for the Tracker layers, and one each for the CZT, CsI, and ACD. Fig.~\ref{fig:triggerModule} shows the trigger module in its housing. Each subsystem produces a fast trigger signal with the Tracker's layers acting independently from each other. If an interaction occurs in a subsystem, it will send a trigger to the TM, which will check for coincidence between other subsystems and flag it appropriately. If the triggers pass the coincidence logic, a trigger acknowledgment along with an Event ID is distributed to all the subsystems. The TM is programmable to enable 16 coincidence logic modes with various permutations of the different subsystems.

The ROSSPAD, which is the front-end readout for the CsI calorimeter and the ACD, does not have the general-purpose input/output capability to accept the Event ID from the TM. Therefore, we utilize an Arduino Due to accept EventIDs and synchronize them with the ROSSPAD triggers. The system also accepts a pulse-per-second (PPS) signal that is converted with a TTL-LVDS board to maintain synchronization between the different components. The Arduino and the TTL-LVDS converter boards are shown in Fig.~\ref{fig:subsystem_CSI}b.

\section{Preliminary Performance}
\label{sec:performance}

This section presents the performance of individual subsystems as well as the performance of the integrated system.

\subsection{Individual System Performance}

Before full system integration, each subsystem has individually undergone testing and calibrations to validate performance. The Tracker underwent a series of inspections to study the uniformity, depletion voltage, and inter-strip capacitance of the DSSDs. 

\begin{figure}[h!]
\includegraphics[width=0.6\linewidth]{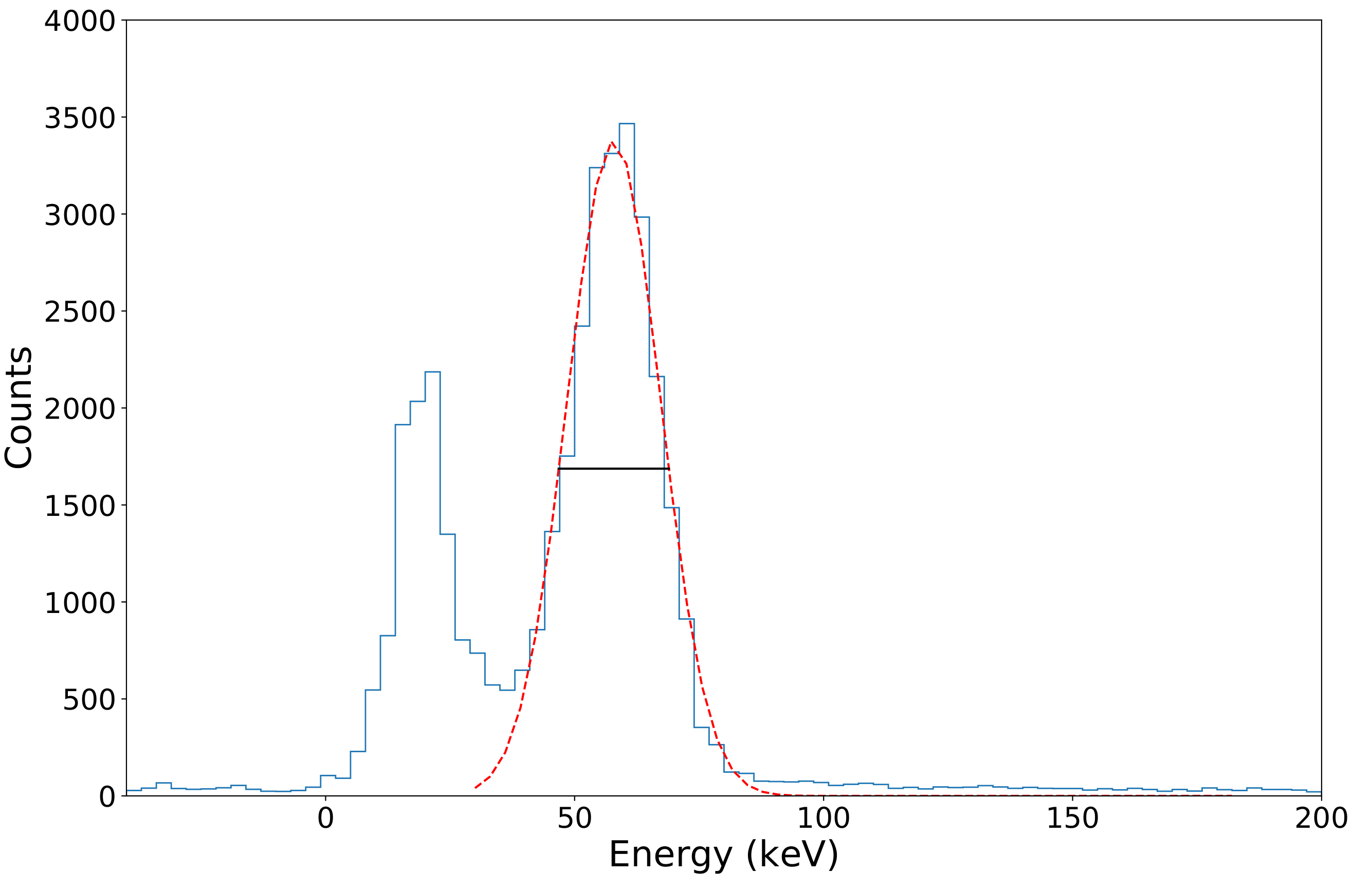}
  \centering
  \caption{Americium 241 spectrum as recorded by the live strips on a single Tracker layer. Note the second peak left of the americium peak is an artifact of the ASIC's pedestal.}
  \label{fig:trkAm241}
\end{figure}

At the time of writing, six of the ten layers have been assembled and characterized. Fig.~\ref{fig:trkAm241} presents an Am-241 spectrum from all the live strips in layer 2 with the $59.5 \ \mathrm{keV}$ line measured with a full-width-at-half-maximum (FWHM) of $22.1 \ \mathrm{keV}$, or $37.1\%$. The energy range of each Tracker strip is estimated to be around $40-700 \ \mathrm{keV}$.

\begin{figure}[h]
\includegraphics[width=0.6\linewidth]{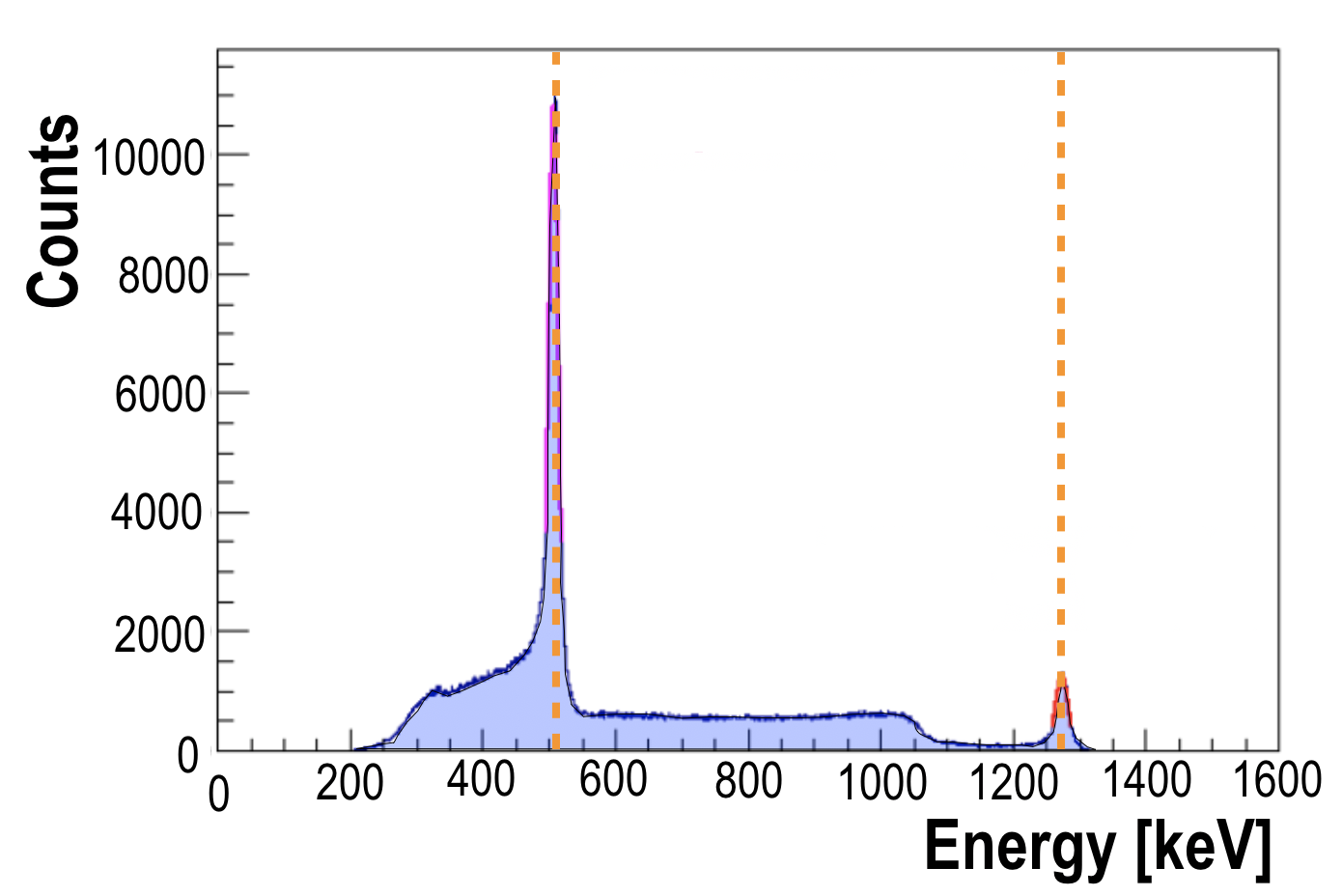}
  \centering
  \caption{CZT spectrum of a Na-22 check source. Note that the $0.511 \ \mathrm{MeV}$ line shows Doppler broadening.}
  \label{fig:cztNa22}
\end{figure}

The CZT subsystem has undergone rigorous testing with its performance reported numerous times~\cite{4x4CZT, bnlAVG, bnlCZT}. Fig.~\ref{fig:cztNa22} presents the CZT's response to an Na-22 source showing $1.8 \%$ FWHM resolution at the $1.274 \ \mathrm{MeV}$. The energy range of the CZT subsystem is estimated to be 0.2-10 MeV per bar.

\begin{figure}[h]
    \centering
    \subfloat[5 MeV]{%
    \includegraphics[height=0.21\textheight]{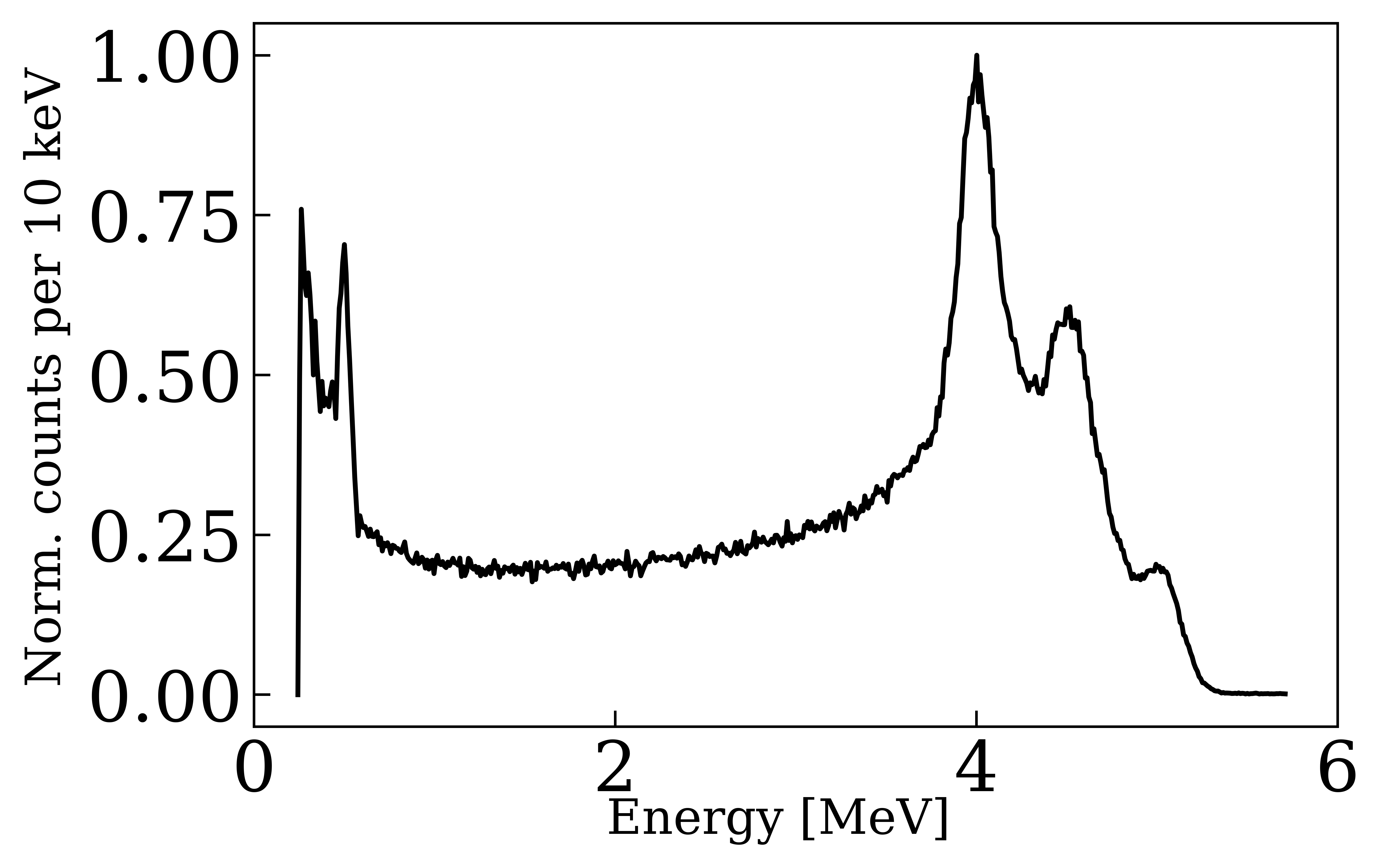}}
    \hspace*{1 cm}
    \subfloat[Muon Response]{%
    \includegraphics[height=0.21\textheight]{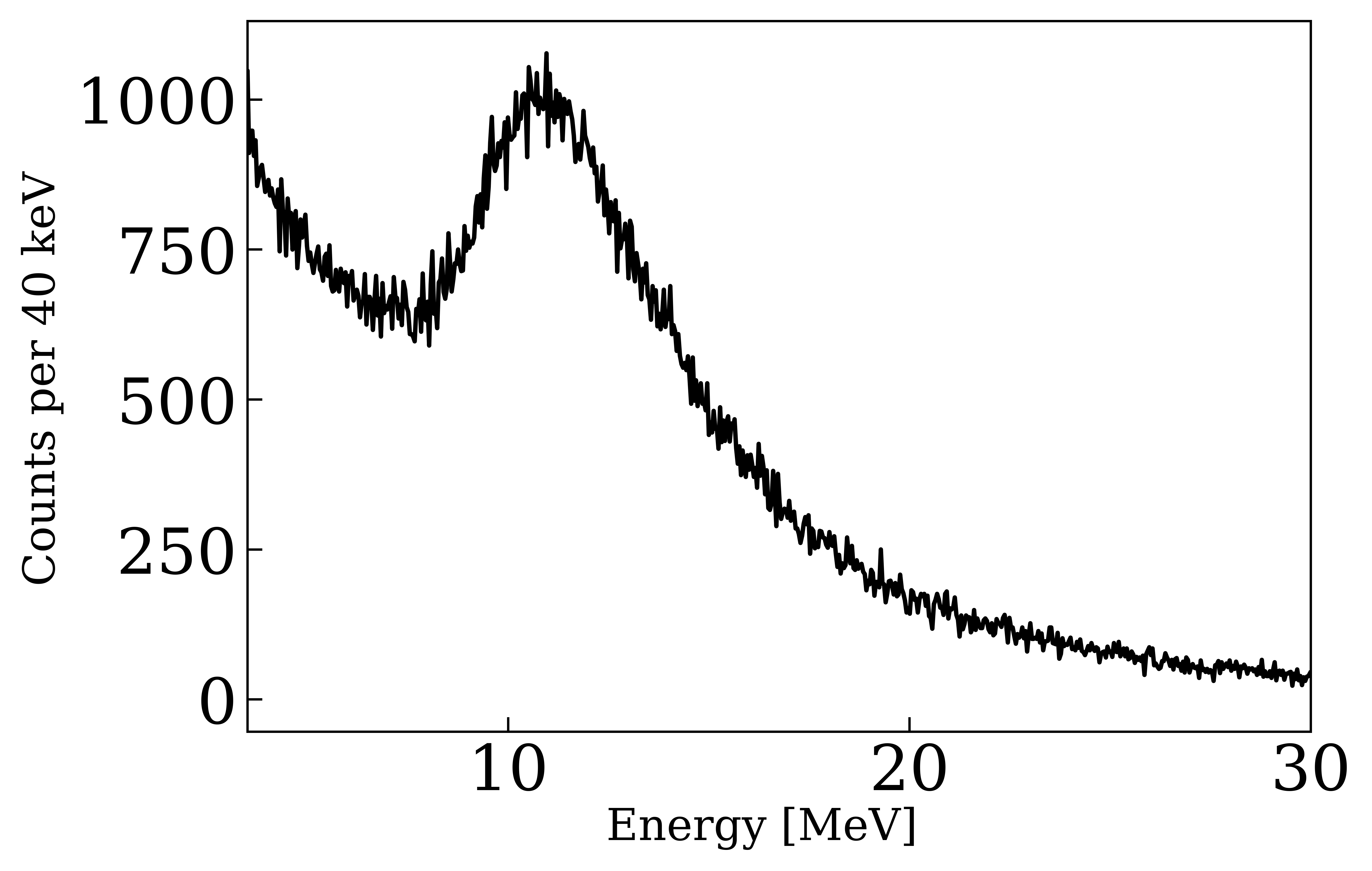}}
\caption{(a) 5 MeV beam spectral response of single log from the CsI calorimeter spectral response. The $5 \ \mathrm{MeV}$ full energy peak is clearly seen along with its escape peaks. At the lower end, the $0.511 \ \mathrm{MeV}$ annihilation peak is visible. (b) The CsI calorimeter's response to muons at around sea level using all channels. Note that (b) treats each interaction in the calorimeter as an independent event.}
\label{fig:csiPerformance}
\end{figure}

We tested the CsI subsystem extensively in the lab with various check sources at energies such as $0.511 \ \mathrm{MeV}$ (Na-22) and $2.6 \ \mathrm{MeV}$ from Th-228. The energy resolution across all individual bars in the system averages around $8.0 \%$ FWHM at $662 \ \mathrm{keV}$. The energy range of a single bar is set to be $0.25 \minus 30 \ \mathrm{MeV}$. In addition, the CsI calorimeter has undergone various tests from monoenergetic beams. Fig.~\ref{fig:csiPerformance}a plots the energy response of a single bar to a monoenergetic $5 \ \mathrm{MeV}$ gamma-ray beam. Fig.~\ref{fig:csiPerformance}b plots the calorimeter's response to muons showing a clear Landau distribution.

\subsection{Fully Integrated System}

\begin{figure}[h]
\includegraphics[trim={0cm 0cm 0cm 0cm},clip, width=0.4\linewidth]{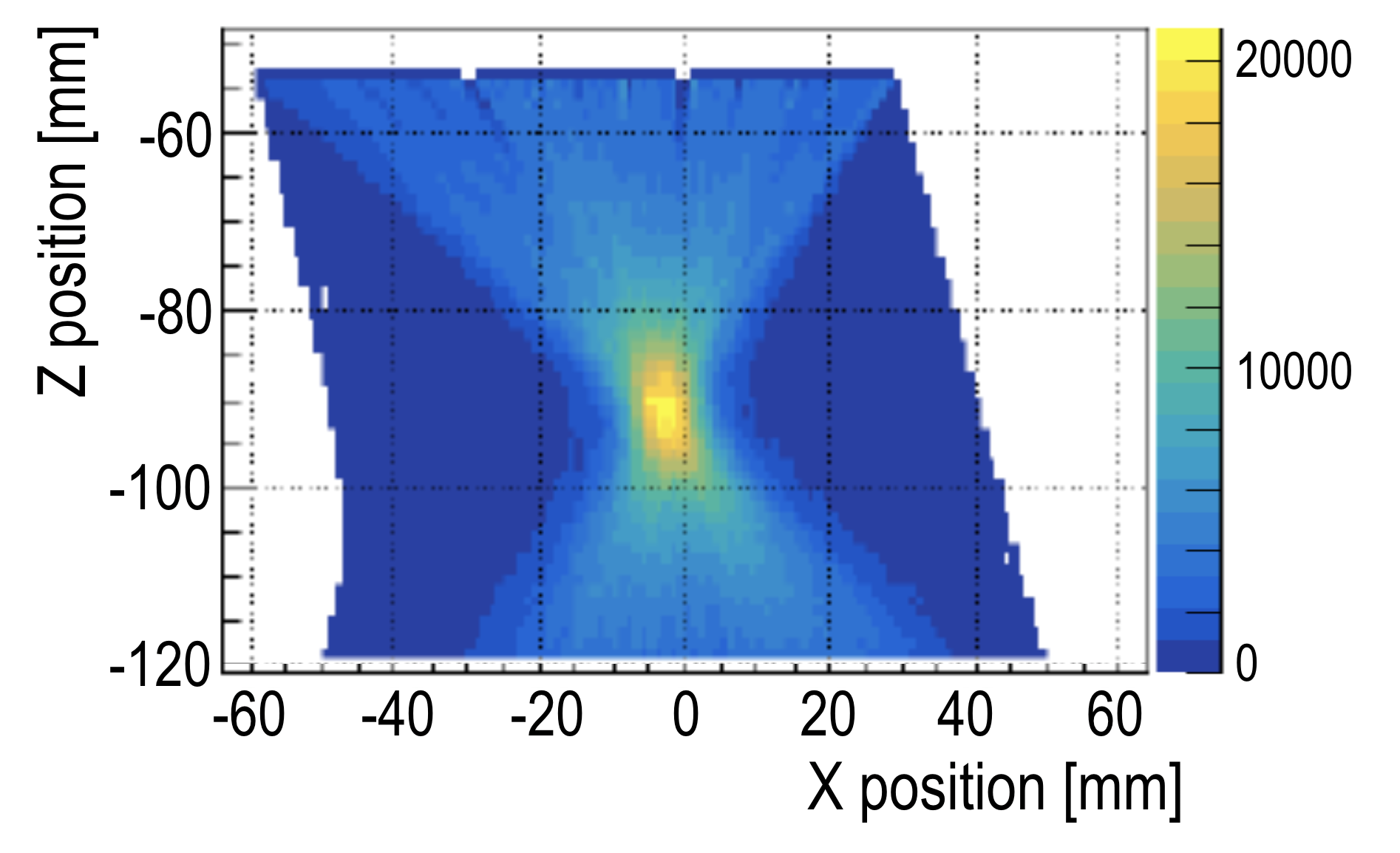}
  \centering
  \caption{A tomographic image plotting the lines of responses of Na-22 annihilation lines when the source is placed between the CZT and CsI. This is essentially a positron emission tomography image and only considers the top layer of the CsI instrument.}
  \label{fig:petCompair}
\end{figure}

We conducted several tests and experiments to validate the integrated system during the instrument commissioning phase. To test the timing and position calibration of the CZT and CsI subsystem, a Na-22 check source was placed in between the active area of the two systems. As the Na-22 source undergoes $\beta^+$ decay, it emits two characteristic $0.511 \ \mathrm{MeV}$ annihilation photons in anti-parallel directions. Simultaneous detection by the CZT and CsI subsystems of the $0.511 \ \mathrm{MeV}$ photons indicates that the triggering mechanism is actively working and assures that the geometry is appropriately modeled. Fig.~\ref{fig:petCompair} plots the lines of response between the locations of the detected $0.511 \ \mathrm{MeV}$ photons. Since a clear hot spot is visible, this implies a simultaneous detection of the $0.511 \ \mathrm{MeV}$ photons and a verification of coincidence. The hot spot also indicates that position reconstruction is functioning.

\begin{figure}[h]
    \centering
    \subfloat[Muon 1]{%
    \includegraphics[height=0.28\textheight]{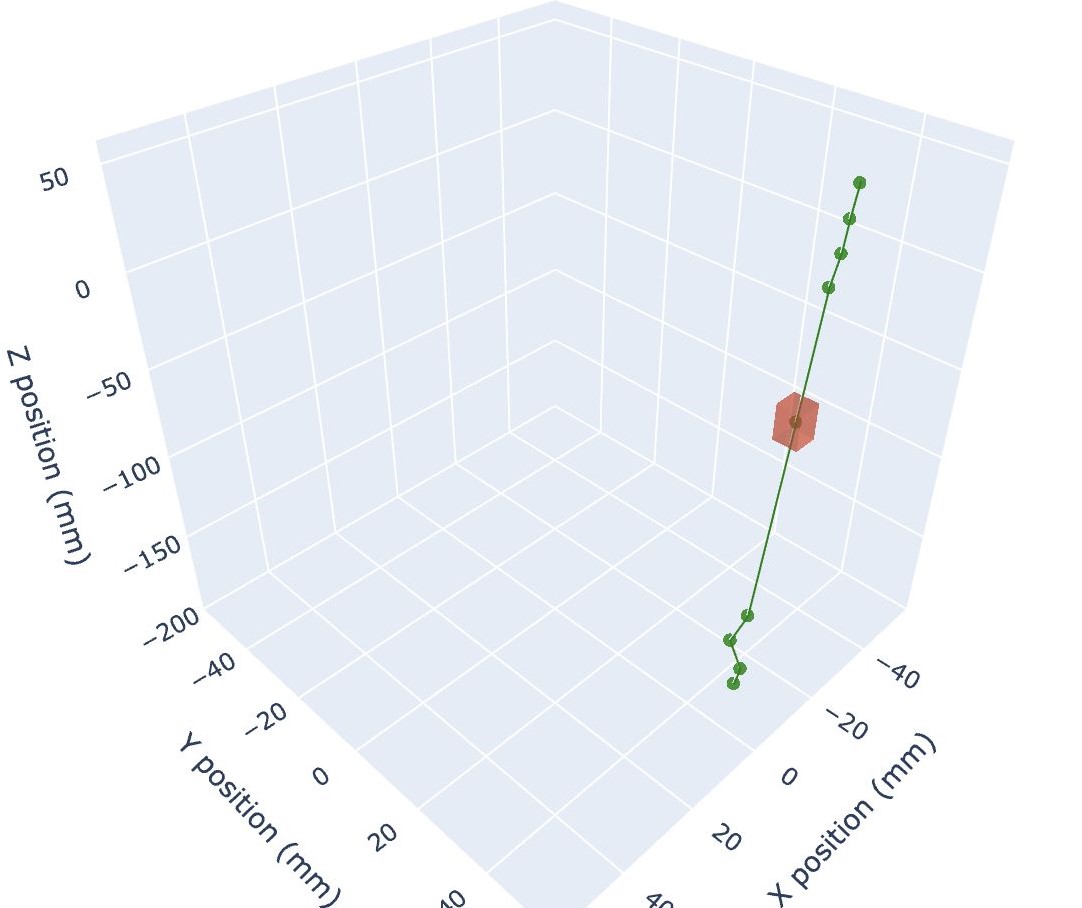}}
    \hspace*{0.8 cm}
    \subfloat[Muon 2]{%
    \includegraphics[height=0.28\textheight]{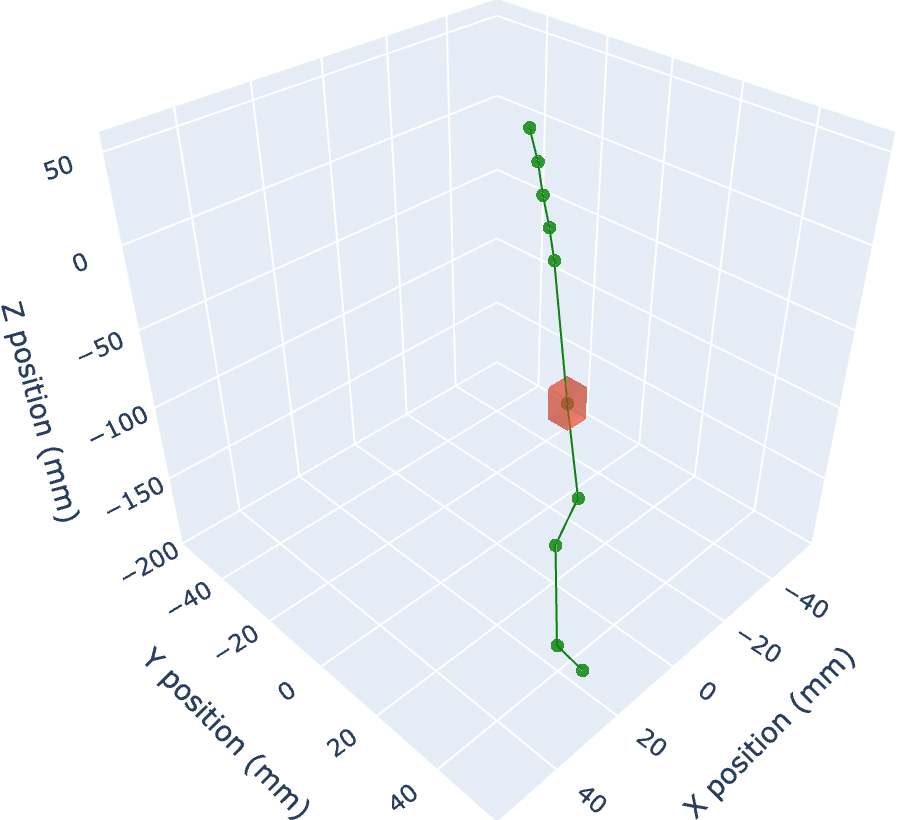}}
\caption{Muons tracks in the core ComPair instrument. Note that the position resolution of the CsI calorimeter is $1 \ \mathrm{cm}$ FWHM. The top set of dots represent hits seen by the Tracker while the middle red cube represents the error boundaries of the hit in CZT. The bottom set of points represent hits in the CsI calorimeter. }
\label{fig:compairMuons}
\end{figure}

When the Tracker was incorporated, cosmic rays presented a convenient tool to validate its integration~\cite{glastBalloon}. Muons are an attractive source in which to verify the coincident triggering mechanism and the geometry of the system due to their greater mass and electric charge. They are therefore very penetrating and travel in straight lines. Fig.~\ref{fig:compairMuons} plots two different muon tracks traversing ComPair. In the plot, the positive Z represents the top of the instrument, where the Tracker is located. The red box represents the error associated with the position reconstructed by the CZT. No error considerations or visual aids are added to the CsI calorimeter tracks. 

The linear tracks indicate the coincidence logic that requires a trigger among all subsystems. Next, since the muons travel in a straight line, recording such a behavior is another indication that integration is successful.

\subsection{Preliminary results from HIGS Beam Test 2022}

In April 2022, a beam test of the prototype without the ACD was performed at Triangle Universities Nuclear Laboratory's High Intensity Gamma-ray Source (HIGS) facility, which delivers monoenergetic gamma-ray beams. The energies studied include 2, 5, 7, 15, and 25 MeV. We present the preliminary results here.

\begin{figure}[h]
\includegraphics[trim={0cm 0cm 0cm 0cm},clip, width=0.5\linewidth]{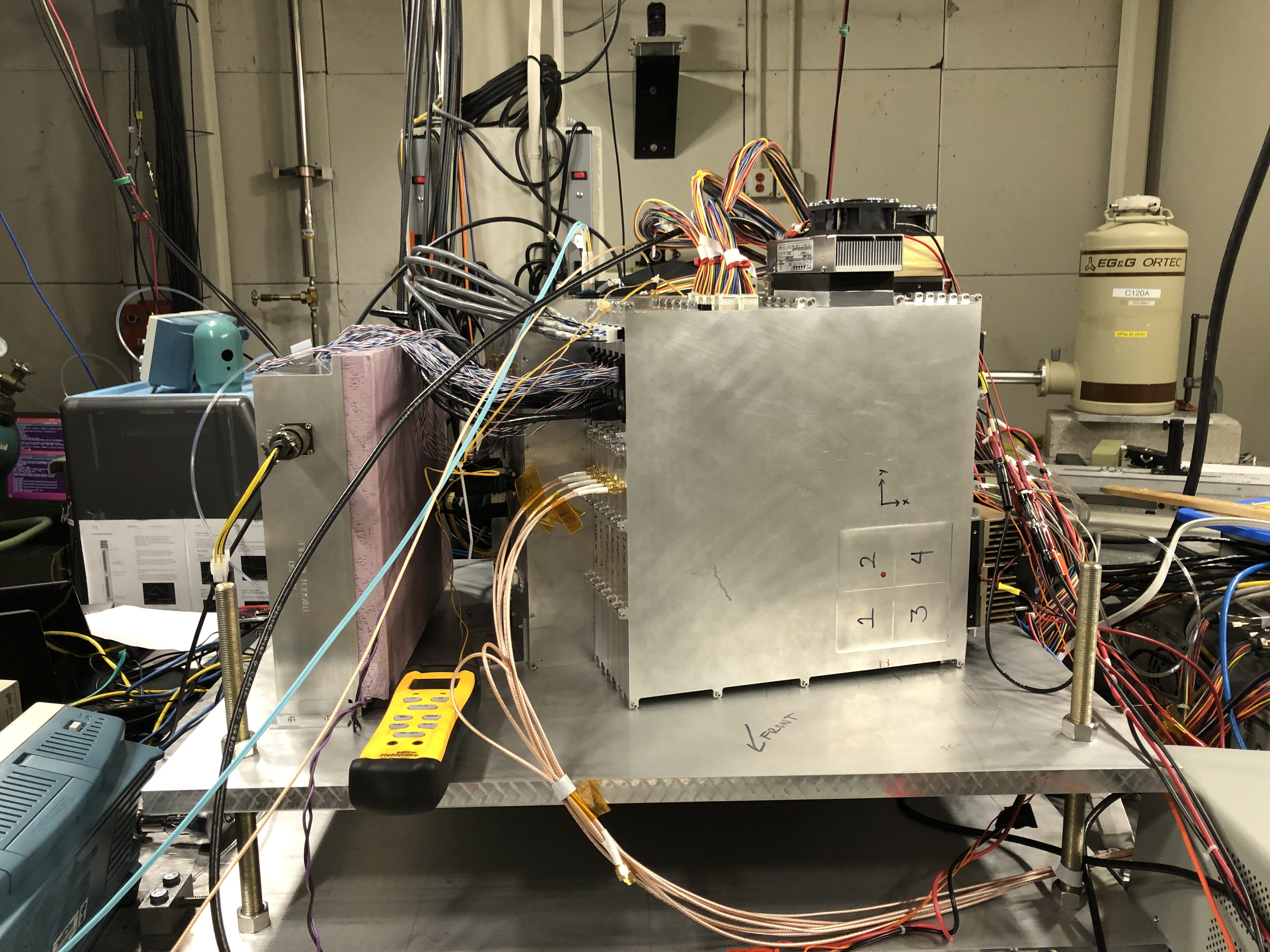}
  \centering
  \caption{Experimental setup of the ComPair system at the HIGS facility. In this configuration, the system is offset 15 degrees with respect to the beam. The beam's vector is into the page aimed at the leftmost side of the DSSD active area. The non-offset configuration has the beam aimed towards the center of the CZT array.}
  \label{fig:compairHIGSSetup}
\end{figure}

\begin{figure}[h]
\includegraphics[trim={0cm 0cm 0cm 0cm},clip, width=0.75\linewidth]{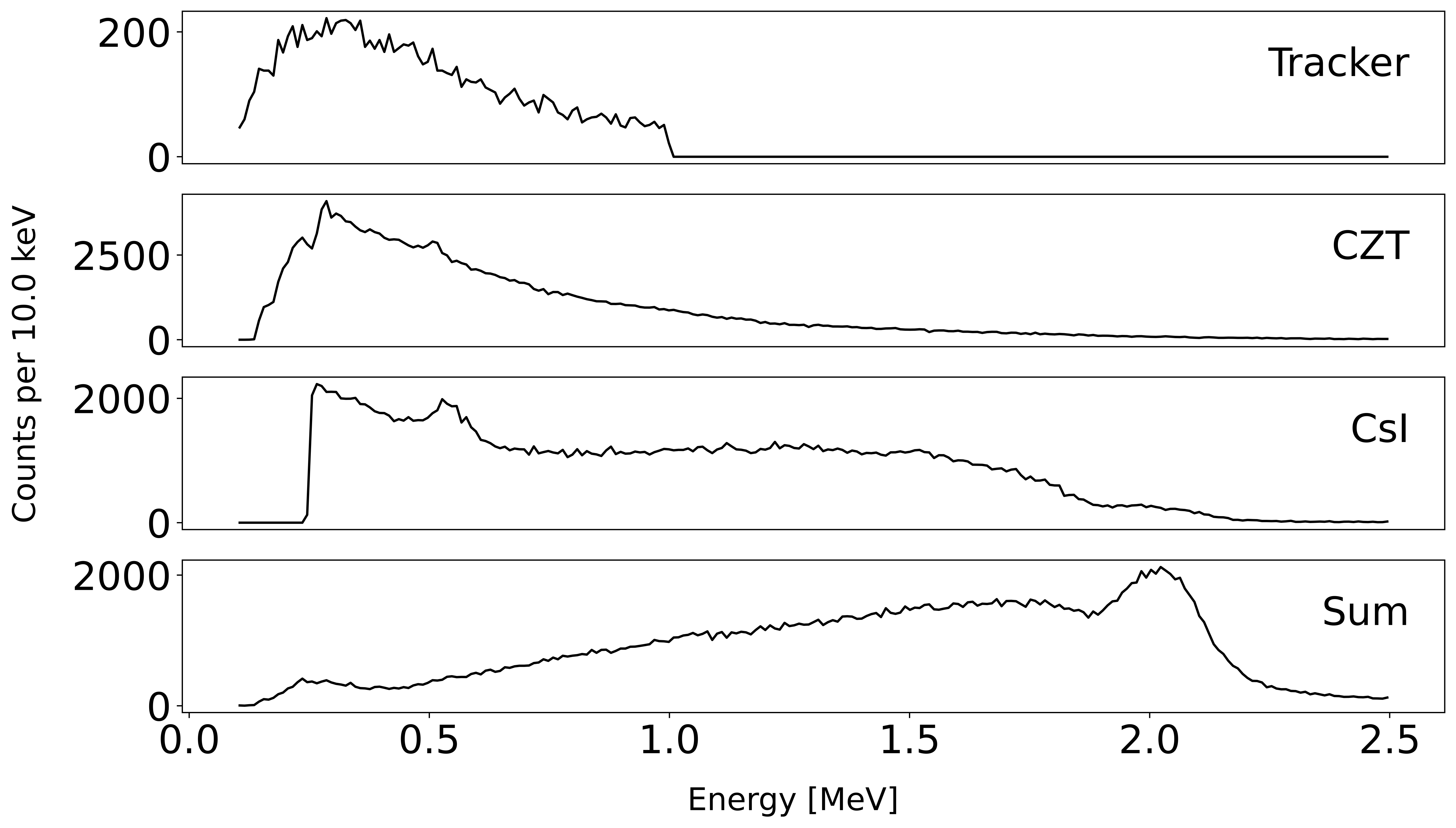}
  \centering
  \caption{Spectral responses of each subsystem as well as the full instrument response to monoenergetic 2 MeV photons. }
  \label{fig:compairHIGS2MeV}
\end{figure}

Fig.~\ref{fig:compairHIGSSetup} shows the experimental setup at the HIGS facility. During the runs, the implemented coincidence allows for full-system readout after any two subsystems have triggered. Fig.~\ref{fig:compairHIGS2MeV} presents the spectral responses to a monoenergetic beam for each subsystem and the summed energies  across all subsystems, showing a full-energy peak at 2 MeV. Note that the results are preliminary as more advanced calibrations are currently under investigation.

\begin{figure}[h]
    \centering
    \subfloat[Experiment]{%
    \includegraphics[height=0.28\textheight]{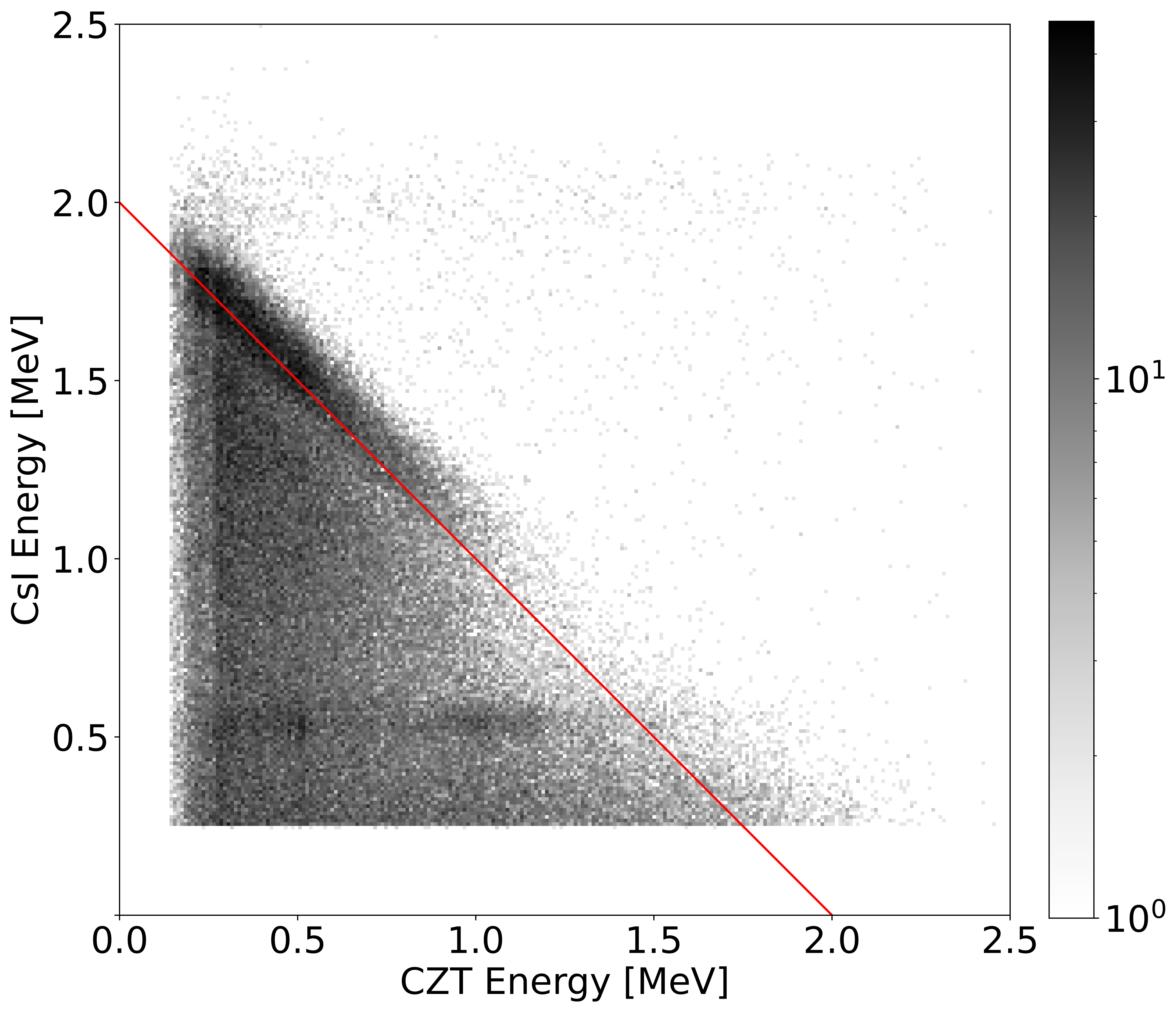}}
    \hspace*{0.8 cm}
    \subfloat[Simulation]{%
    \includegraphics[height=0.28\textheight]{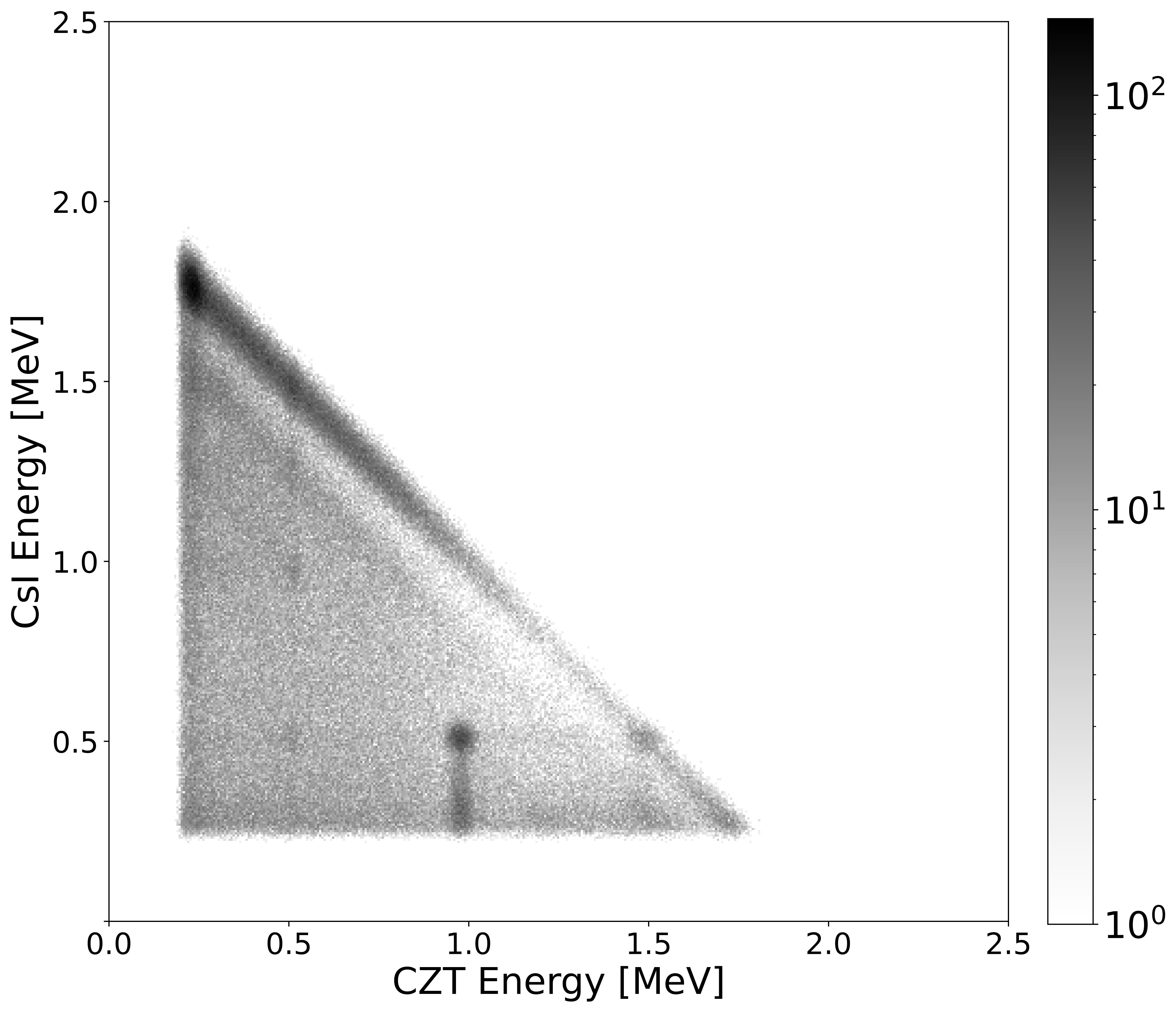}}
\caption{Experimental and simulated bivariate histograms of energy responses in the CZT and CsI calorimeters. The 1 MeV band in the CZT represents the double-escape peak with one of the annihilation photons detected in the CsI. The 1.5 MeV band in CZT represents the single escape peak with a single annihilation photon in the CsI. The diagonal of the lower triangle represents the fully contained gamma ray in the two subsystem. The island in the diagonal in the CZT energies between $0.25-1 \ \mathrm{MeV}$ range represent scatter in the CZT and a full energy deposition in the CsI.}
\label{fig:simExpComparison}
\end{figure}

To study the energy responses of the subsystem, Fig.~\ref{fig:simExpComparison} plots the bivariate nature of the energies deposited in the CsI and CZT calorimeters for (a) experimental and (b) simulated data. Several features are visible in both experiment and simulation, mainly the double escape peak in the CZT and an escaped annihilating photon detected in the CsI. In addition, small angle scatters in the CZT with a photo-absorption in the CsI region are clearly visible in both the experiment and simulation.

\section{Future Work}
\label{sec:futureWork}

\begin{figure}[h]
    \centering
    \subfloat[Instruments block diagram]{%
    \includegraphics[height=0.2\textheight]{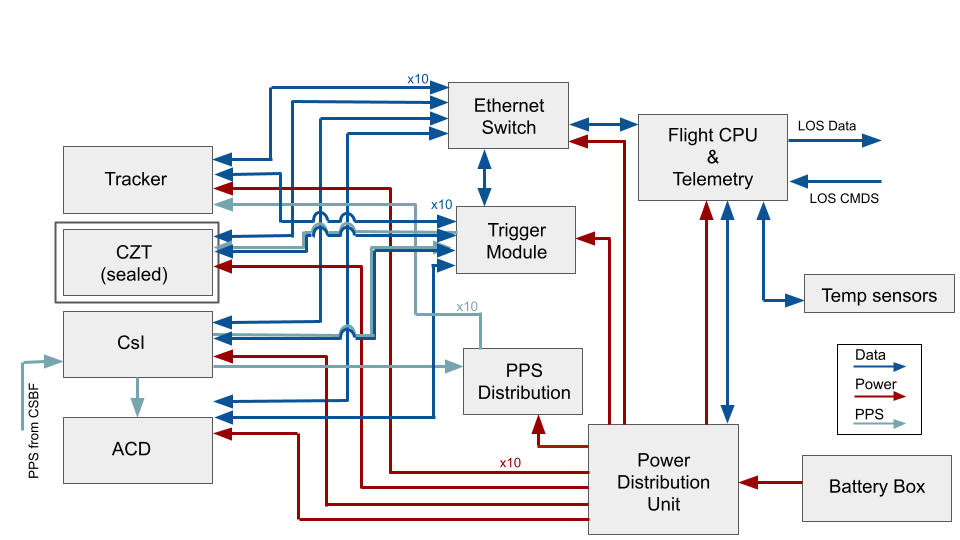}}
    \hspace*{0.8 cm}
    \subfloat[Balloon Instrument]{%
    \includegraphics[height=0.2\textheight]{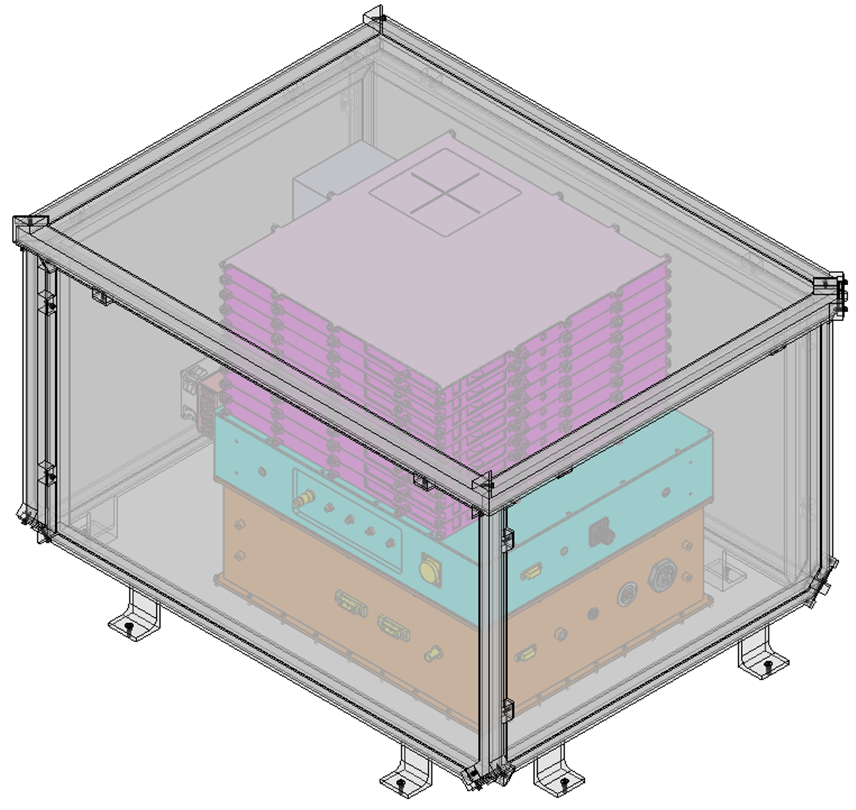}}
\caption{(a) Block diagram of the instrument detailing data flow, power and tertiary systems. (b) CAD of the balloon instrument.}
\label{fig:balloon}
\end{figure}

Analysis of HIGS beam test is underway with more results expected in the future. Next, in fall of 2023, the system is slated to fly as a short-duration balloon experiment out of Fort Sumner, NM, hosted by NASA's Columbia Scientific Balloon Facility (CSBP). The system is therefore undergoing design upgrades to a balloon flight-ready prototype. This includes upgrading the CZT housing to a hermetically sealed container to maintain appropriate high-voltage safety and ensure high performance. The upgrade also includes the integration of the ACD. Finally, each subsystem will undergo thermal-vacuum testing to characterize and study its performance. Fig.~\ref{fig:balloon}a presents the instrument's block diagram with Fig.~\ref{fig:balloon}b presenting the notional balloon instrument.

\subsection{Flight Computer}
The ComPair DAQ computer is under development by the Los Alamos National Laboratory (LANL) and acts as the data acquisition CPU for the different subsystems. As the main controller, it will send appropriate commands to each subsystem and store data during the flight. Its architecture is based on a Versalogic BayCat single board computer. 

During the balloon flight, the computer will supervise subsystem-specific long-running processes, manage data collection, control the power distribution unit and readout thermometry. The instrument will also connect to the CSBP miniature science instrument package to telemeter housekeeping and a subset of data during the flight.

\subsection{Further Technology Development}

Several technologies are currently in development for future observatories. The Tracker group is looking into implementing Si Pixel technology~\cite{astroPix}. The CZT group is developing a digital system that can digitize the waveforms that yield better energy and position resolution. The CsI group is developing Dual-Gain SiPMs~\cite{dualGainNSS} to extend the dynamic range while maintaining energy linearity with high position and energy resolution. Also in development is a `finger' array rather than a hodoscopic geometry which has vertical short chunks placed in an array and readout from the top and bottom, rather than the sides. This has the potential of delivering higher position resolution. Improvements to the image processing code, MEGALib, is being done by advancing pair reconstruction techniques and optimization for ComPair and AMEGO. This includes using single-site events to improve the detection and localization efficiency of gamma sources~\cite{AMEGO_singles}.

\section{Conclusion}
\label{sec:misc}

The ComPair instrument has been developed as a prototype for AMEGO to demonstrate its technological potential for a MeV gamma ray observatory. Each subsystem and the integrated system as a whole have undergone testing and verification. An augmentation to AMEGO, named AMEGO-X~\cite{amegox}, which uses Silicon pixels and a CsI calorimeter, is currently proposed as a medium-explorer class instrument.  The technology developed in ComPair directly supports AMEGO-X and the training of early career scientists in MeV gamma-ray instrument development. 

%\end{comment}

\acknowledgments % equivalent to \section*{ACKNOWLEDGMENTS}       
 
This work is supported under NASA Astrophysics Research and Analysis (APRA) grants NNH14ZDA001N-APRA, NNH15ZDA001N-APRA, NNH18ZDA001N-APRA, NNH21ZDA001N-APRA.

% References
\bibliography{report} % bibliography data in report.bib
\bibliographystyle{spiebib} % makes bibtex use spiebib.bst

\end{document}